\documentclass{article}

\usepackage{PRIMEarxiv}

\usepackage[utf8]{inputenc} 
\usepackage[T1]{fontenc}    
\usepackage{url}            
\usepackage{booktabs}       
\usepackage{amsfonts}       
\usepackage{amsmath}
\usepackage{amssymb}
\usepackage{nicefrac}       
\usepackage{microtype}      
\usepackage{lipsum}
\usepackage{lastpage,fancyhdr,graphicx}

\graphicspath{{media/}}     
\usepackage{authblk}

\usepackage{nameref,hyperref}
\usepackage[right]{lineno}
\DisableLigatures[f]{encoding = *, family = * }

\usepackage[english]{babel}
\usepackage{mathtools}
\usepackage{subcaption}
\usepackage{float}
\usepackage[ruled,vlined]{algorithm2e}
\usepackage[backend=biber,style=nature,citestyle=nature]{biblatex}
\usepackage{csquotes}
\usepackage[symbol]{footmisc}

\usepackage{epstopdf}
\usepackage{parskip}
\usepackage{color}

\definecolor{Gray}{gray}{.25}

\usepackage{sidecap}

\usepackage{wrapfig}
\usepackage[pscoord]{eso-pic}
\usepackage[fulladjust]{marginnote}
\reversemarginpar

\usepackage{dsfont}
\usepackage{commath}
\usepackage{bbm}
\usepackage{amsmath, bm}
\usepackage{listings}
\usepackage{glossaries}

\usepackage{setspace}
\raggedright
\usepackage{changepage}

\usepackage[aboveskip=1pt,labelfont=bf,labelsep=period,singlelinecheck=off]{caption}
\makeatletter
\renewcommand{\@biblabel}[1]{\quad#1.}
\makeatother

\title{On the distribution of cosine similarity with application to biology}

\author[1,2]{Ian Smith}
\author[6, 7, 8]{Janosch Ortmann}
\author[1]{Farnoosh Abbas-Aghababazadeh}
\author[1]{Petr Smirnov}
\author[1,2,3,4,5*]{Benjamin Haibe-Kains}
\affil[1]{Princess Margaret Cancer Centre, University Health Network, Toronto, Canada}
\affil[2]{Department of Medical Biophysics, University of Toronto, Toronto, Canada}
\affil[3]{Ontario Institute for Cancer Research, Toronto, Canada}
\affil[4]{Vector Institute for Artificial Intelligence, Toronto, Canada}
\affil[5]{Department of Computer Science, University of Toronto, Toronto, Canada}
\affil[6]{Département d’analytique, opérations et technologies de l’information, École des sciences de la gestion, Université du Québec à Montréal, Montréal, Canada}
\affil[7]{Groupe d'études et de recherche
en analyse des décisions (GERAD), Monrtréal, Canada}
\affil[8]{Centre de recherches mathématiques (CRM), Montréal, Canada}
\affil[*]{Corresponding author: benjamin.haibe-kains@uhn.ca}
\date{August 2023} 
\begin{document}
\maketitle

\begin{abstract}
    Cosine similarity is an established similarity metric for computing associations on vectors, and it is commonly used to identify related samples from biological perturbational data. The distribution of cosine similarity changes with the covariance of the data, and this in turn affects the statistical power to identify related signals. The relationship between the mean and covariance of the distribution of the data and the distribution of cosine similarity is poorly understood. In this work, we derive the asymptotic moments of cosine similarity as a function of the data and identify the criteria of the data covariance matrix that minimize the variance of cosine similarity. We find that the variance of cosine similarity is minimized when the eigenvalues of the covariance matrix are equal for centered data. One immediate application of this work is characterizing the null distribution of cosine similarity over a dataset with non-zero covariance structure. Furthermore, this result can be used to optimize over a set transformations or representations on a dataset to maximize power, recall, or other discriminative metrics, with direct application to noisy biological data. While we consider the specific biological domain of perturbational data analysis, our result has potential application for any use of cosine similarity or Pearson’s correlation on data with covariance structure.
\end{abstract}

\section*{Introduction}
Many biological datasets have analysis problems that require computing similarities between elements of the dataset. Perturbational datasets, where changes in a feature space induced by a perturbation or disease are measured, are one such application where relationships between perturbations, reagents, and diseases are modeled. For instance, in the L1000 Connectivity Map\cite{Subramanian2017} or the Cell Painting morphological assay\cite{bray_cell_2016}, cancer cell lines are treated with small molecules (drugs) and genetic reagents, and signatures of induced biological changes are measured in gene expression and cellular morphological features respectively. A typical analysis involves applying a similarity function to pairs of signatures - vectors of differential features - to identify similar and dissimilar reagents to identify mechanism of action, to find gene agonists or antagonists, and to build biological networks. 

An important challenge with perturbational data is the low signal-to-noise ratio: perturbations produce small reproducible changes compared to the scale of the biological and measurement noise. This is a consequence of homeostasis inducing stability in a dynamical system: sufficiently large changes have catastrophic outcomes that are not tolerated by cells \cite{nijhout_systems_2019, tyson_dynamical_2020}. Low signal-to-noise is exacerbated by the curse of dimensionality, where changes in a small number of features are obscured by noise in the high number of dimensions of the space \cite{pestov_geometry_2000}. Increasing this signal-to-noise ratio, or the ability to discriminate pairs of similar signatures from the background maximizes the usefulness of perturbational datasets to identify related compounds, find biomarkers, and discover biological relationships. Embedding the data in a new representation or transformation of the feature space is a purely computational strategy to achieve this, overcome limitations of the assay, and discover new relationships.

Biological feature spaces are generally anisotropic manifolds with complex covariance structure \cite{chen_estimating_2018, moon_manifold_2018, boileau_exploring_2020}. We seek to characterize the distribution of the cosine similarity on the data with the aim of minimizing the variance to maximize discriminative power. Finding transformations on the data to maximize power makes data more useful for identifying similar data points - for perturbational data, characterizing drug mechanism of action, grouping gene inhibitions into pathways, and nominating therapeutics for disease states. Given a data set, computing the distribution of a similarity function on samples from that space empirically is straightforward, but understanding how transformations applied to the data affect the distribution of a similarity function is unsolved. 

In this work, we characterize the moments of cosine similarity for multivariate distributions. We consider multivariate normal as a special case because they are reasonable first approximations of many datasets and because they are mathematically tractable. We find that the variance is minimized for data with mean zero when the eigenvalues of the covariance matrix are equal, and identify under what conditions the variance is minimized for data with non-zero mean. We also connect the distribution of cosine similarity to statistical power for applications like identifying relationships between perturbational signatures. Understanding how changes in the distribution of the eigenvalues of the covariance structure of data affect the distribution of cosine similarity provides a theoretical grounding for representation learning and data transformations to make data sets more useful for discovery. 

This manuscript has been organized to maximize clarity. First, we define the problem and present the mathematical findings for three cases of data distribution. Second, we tie the results to discriminative power, check the findings with empirical simulations, and present discussion to contextualize the result. The detailed derivations of all the findings are presented in the derivations section after the discussion. Finally, for convenience, an appendix section at the end includes some commonly used relationships and characteristic functions. 

\section*{Formal statement of the problem}

Let $X \in \mathbb{R}^{n}$ be a random variable with finite (vector) mean $\boldsymbol{\mu}$, finite covariance matrix $\Sigma$, and probability distribution $f_{x}(X)$. What is the mean and variance of cosine similarity for independent and identically distributed (i.i.d.) elements of X?  

For $A, B \in X$ i.i.d., cosine similarity is defined in the usual way, i.e:
\begin{equation}
\label{eqn:cosine}
\cos(A, B) = \frac{\sum_{i = 1}^{n} A_{i} B_{i}}{\sqrt{\sum_{i} A_{i}^{2}}\sqrt{\sum_{i}B_{i}^{2}}} = \frac{A^{T} B}{\sqrt{(A^{T}A)(B^{T} B)}} = \frac{A^{T}B}{\abs{A}\abs{B}}
\end{equation}

We consider the following three cases. Note that because we can choose the basis such that the covariance matrix is diagonal (Derivations \ref{cosProps}), a finite covariance matrix is equivalent to finite eigenvalues of the covariance matrix. We use the terms 'finite covariance' and 'finite variance' to describe this interchangeably.

\begin{enumerate}
    \item Spherically symmetric case: $X \sim \mathcal{N}(0, \mathbbm{1})$
    \item Mean 0, finite covariance: $\boldsymbol{\mu} = 0$, $\Sigma$ finite.
    \item Finite but non-zero mean and covariance: $\boldsymbol{\mu}$, $\Sigma$. 
\end{enumerate}

\subsection*{Conventions}

For simplicity, we adopt a few conventions throughout this manuscript.
\begin{itemize}
    \item Unless otherwise noted, summations have a range from 1 to n. $\sum_{k} := \sum_{k=1}^{n}$
    \item We denote the covariance matrix of $X \in \mathbb{R}^{n}$ as $\Sigma$, and the ith eigenvalue of $\Sigma$ as $\sigma_{i}^{2}$.
    \item Capital letters are used to denote vector random variables of length n, e.g. $A, B, X$, unless otherwise noted. The ith scalar element of a vector $A$ is $A_{i}$. For instance, $A = [A_{1}, A_{2}, A_{3}, ... , A_{n}]$. 
    \item Vector-valued parameters, like the mean $\boldsymbol{\mu} = [\mu_{1}, \mu_{2}, ..., \mu_{n}]$ are denoted in bold. $\mu_{i} \equiv \mathbb{E}[A_{i}]$.
    \item Scalar-valued parameters, like the dimension n, are denoted by lowercase letters. 
\end{itemize}
Abbreviations:
\begin{itemize}
\item IID / i.i.d. - independent and identically distributed
\item WLOG / w.l.o.g. - without loss of generality
\item $\mathbb{I}_{N}$ - the NxN identity matrix  
\end{itemize}

\section*{Case 1: Spherical symmetry: multivariate standard normal}

The case where $X \sim \mathcal{N}(\boldsymbol{0}, \mathbbm{1})$, with an identity covariance matrix and dimension $n$, is straightforward to calculate. The components of $X$ are independent and identically distributed standard normal random variables, but because of the rotational and scale invariance of cosine similarity, this is equivalent to the case where all the eigenvalues of the covariance matrix are equal (Derivations \ref{cosProps}). As shown in Muller and Marsaglia, $X/\abs{X}$ is a uniform probability distribution over the unit sphere, $S^{n-1}$ \cite{muller_note_1959, marsaglia_choosing_1972}. 

The spherical symmetry of $\mathcal{N}(\mathbf{0}, \mathbbm{1})$ can be exploited, because the probability distribution of cosine is the same as the marginal distribution of a single component of $X/\abs{X}$. That is, one of the arguments to cosine can be chosen without loss of generality (WLOG) to be $A = (1, 0, 0, ...)$. For the full derivation, see Derivations section \ref{case1Deriv}. For $B \sim X$, this gives:
\begin{equation}
\frac{1 + \cos(A,B)}{2} \sim Beta(\frac{n-1}{2}, \frac{n-1}{2})
\end{equation}
\begin{align}
    \mathbb{E}[\cos(A,B)] &= 0 \\
    Var(\cos(A,B)) &= \frac{1}{n}
    \label{eqn:Case1}
\end{align}
\section*{Case 2: Centered distribution with unequal eigenvalues}

We next consider the more general problem of a mean zero distribution with unequal but finite variances. Because cosine is invariant under coordinate rotation with an orthogonal matrix, the covariance matrix can WLOG be assumed to be diagonal (Derivations \ref{cosProps}). Furthermore, all the eigenvalues of the covariance matrix can be assumed to be non-zero, as an $n$ dimensional space with $k$ eigenvalues equal to zero reduces to an $n-k$ dimensional subspace with non-zero eigenvalues. Therefore, let the components $X_{i}$ be independent but not identically distributed: $X_{i} \sim f_{X_{i}}(x_{i})$ with mean 0 and variance $\sigma_{i}^{2}$. The eigenvalues $\lambda_{i}$ of the covariance matrix are $\lambda_{i} = \sigma_{i}^{2}$. Let $A,B \sim X$ i.i.d.

A closed form or even the moments of the distribution of $\cos(A,B)$ from Equation \ref{eqn:cosine} are difficult to evaluate because the expression is a ratio of random variables. See Derivations \ref{case2Deriv} for more detail with the specific case of $X \sim \mathcal{N}(\mathbf{0}, \Sigma)$. 

However, we can approximate the denominator, which simplifies the calculation. Using cosine's scale invariance property, for $m = \sqrt{\sum_{i} \sigma_{i}^{2}}$, define $b(X)$ as:

\begin{equation}
    b(X) \equiv \frac{X}{\sqrt{\sum_{i} \sigma_{i}^{2}}} = \frac{1}{m} X
    \label{normapprox}
\end{equation}

By design, the square of the norm of the rescaled random variable, $|b(X)|^{2}$, has expectation 1. The ratio of the variance to the square of the mean is $\frac{\sum_{k}2\sigma_{k}^{2}}{(\sum_{k}\sigma_{k}^{2})^{2}}$ and tends to 0 in the limit as the dimension $n \rightarrow \infty$ provided the contribution of any one covariance eigenvalue is sufficiently small (Derivations \ref{normMultivar}). This means that we can approximate the norm of a multivariate vector as defined as a constant equal to the square root of the expectation of the squared norm. For sufficiently large n, we can then approximate $\cos(A,B)$:

\begin{align}
    \cos(A,B) &= \cos(b(A), b(B)) = \frac{\sum_{i}b(A)_{i}b(B)_{i}}{|b(A)||b(B)|} \\
    &\approx \sum_{i}b(A)_{i}b(B)_{i}
\end{align}

Using this approximation, we can compute the moments of $\cos(A,B)$:
\begin{align}
    \mathbb{E}[\cos(A,B)] &= 0 \\
    Var(\cos(A, B)) &= \sum_{i} \frac{\sigma_{i}^{4}}{(\sum_{j} \sigma_{j}^{2})^{2}}
    \label{eqn:Case2}
\end{align}

Taking the gradient of the approximation with respect to the component variances $\sigma_{i}$ yields:
\begin{equation}
    \frac{\partial Var(\cos(A, B))}{\partial \sigma_{i}^{2}} = (\sum_{j} \sigma_{j}^{2}) \sigma_{i}^{2} - (\sum_{j} \sigma_{j}^{4})
\end{equation}

The global minimum of the approximation of $Var(\cos(A, B))$ is then achieved when $\sigma_{i}^{2} = \sigma_{j}^{2} \forall i, j$, i.e. when all the eigenvalues of the covariance matrix are equal. We define this as the isotropic principle: that the variance of cosine similarity is minimized with equal eigenvalues for centered multivariate distributions with finite covariance. 

\section*{Case 3: Generalized multivariate distribution}

The third case is a general multivariate distribution with finite mean and variance. As before, we assume WLOG that the covariance matrix is diagonal, and the eigenvalues of the covariance matrix are finite and non-zero. Computing the distribution of cosine similarity is difficult because as with case 2, the cosine of two random variables i.i.d. in $X$ is a random variable that is a ratio of complicated random variables. We can employ the same method as with case 2, where we approximate the norm of the multivariate vectors with the expectation value. With some constraints on the relative scale of the eigenvalues of the covariance matrix and for sufficiently large dimension n, the ratio of the standard deviation to the mean converges to 0 (See \ref{case3Deriv}). As before, let $A,B \sim X$ i.i.d.

Using this approximation, the mean and variance of cosine similarity for vectors i.i.d from $X$ is as follows:
\begin{equation}
    \mathbb{E}[\cos(A, B)] = \frac{\sum_{k} \mu_{k}^{2}}{\sum_{j}\mu_{j}^{2} + \sigma_{j}^{2}}
\end{equation}
\begin{equation}
    Var(\cos(A, B)) = \frac{\sum_{k} \sigma_{k}^{2} (\sigma_{k}^{2} + 2 \mu_{k}^{2})}{(\sum_{j} \mu_{j}^{2} + \sigma_{j}^{2})^{2}}
\end{equation}

Note that the mean and variance of the cosine distribution for Case 3 reduces to the mean and variance for Case 2 when $\mathbf{\mu} = 0$. We also can minimize the variance of $\cos(A, B)$ with respect to the variance of the data $\sigma_{i}^{2}$. However, the minimization of the variance is with respect to transformations on the data that also affect the component means. We can express $Var(\cos(A, B))$ rewriting the means as dimensionless factors $\eta$ in units of the variance:
\begin{equation}
    \mu_{i} = \eta_{i} \sigma_{i}
\end{equation}
\begin{equation}
    Var(\cos) = \frac{\sum_{k} \sigma_{k}^{4} (1 + 2 \eta_{k}^{2})}{(\sum_{j} \sigma_{j}^{2} (1 + \eta_{j}^{2}))^{2}}
\end{equation}

The variance of cosine similarity is minimized when the following is true $\forall i, k$ for some constant C:
\begin{equation}
C = \sigma_{i}^{2} \frac{1 + 2 \eta_{i}^{2}}{1 + \eta_{i}^{2}} = \sigma_{k}^{2} \frac{1 + 2 \eta_{k}^{2}}{1 + \eta_{k}^{2}}
\end{equation}
\begin{equation}
    \sigma_{i}^{2} = C \frac{1 + \eta_{i}^{2}}{1 + 2 \eta_{i}^{2}}
\end{equation}
The interpretation is that in the optimal embedding, the ratio of the variances is no greater than 2. The variance of cosine is minimized when component variances with means close to 0 are the largest, and as the mean tends to infinity, component variances tend to 1/2 that of those with mean 0. As a sanity check, when the mean is zero, i.e. $\eta_{i} = 0 \forall i$, the relationship between the variances reduces to the isotropic result from Case 2. 

\subsection*{Modeling discriminative power}

One application of this formulation is computing discriminative power when comparing data from several distributions with cosine similarity. As a practical example, for perturbational datasets \cite{Subramanian2017, bray_dataset_2017, way_morphology_2022, chandrasekaran_jump_2023}, suppose a collection of functionally related compounds C produce signatures from some distribution $P_{C}(X)$ with mean $\mu_{C}$ and covariance $\Sigma$. Suppose there is a collection of all compound signatures $B$ with distribution $P_{B}(X)$ with mean $\mu_{B}$ and the same covariance matrix $\Sigma$. For example, C might consist of ALK inhibitors whose signature is the changes induced by inhibition of ALK, and B could be a large collection of many compound classes, negative controls, and inert agents. The goal of perturbational datasets is to identify compounds that produce similar signatures or biological changes. 

In a typical analysis, for an unknown compound signature $X$, and a set of signatures $S$ for a known class C, the distribution of $\cos(X, S_{i})$ can be compared with the background distribution to determine if $X$ is a member of C. The null hypothesis would be that $X \notin C$ and would be rejected if $\cos(X, S_{i})$ were sufficiently different from the background distribution.

The statistical power is the probability of rejecting the null hypothesis when the alternate is true - in this case, the probability of correctly identifying a compound signature X when $X \in C$. To assess this, we compare the distribution $\cos(S_{a}, S_{b})$ for $S_{a}, S_{b} \in C$, and compare this distribution with the null or background similarity $\cos(B_{a}, B_{b})$ for $B_{a}, B_{b} \in B$.  A chosen statistical level $\alpha$, e.g. $\alpha = 0.01$ determines the threshold $\tau(\alpha)$ as a quantile of the background distribution. The power can be quantified as the probability $P(\cos(S_{a}, S_{b}) > \tau(\alpha)$. There are other appropriate discriminative measures, like signal-to-noise ratio, but power is a useful illustration. 

With the approximation of the mean and variance of $\cos$, we can compute the power by approximating $\cos(X)$ as a normal distribution. Let $Q(\alpha)$ be the quantile function of the normal distribution. The power is then given by the following, where the expression for delta uses the moments for the cosine approximation.
\begin{equation}
    P(\cos(C_{a}, C_{b}) > \tau(\alpha)) = erf\left(\frac{\mathbb{E}[\cos(C_{a}, C_{b})] - \tau(\alpha)}{Var[\cos(C_{a}, C_{b})]^{1/2}}\right) \equiv erf(\Delta)
\end{equation}
\begin{equation}
    \Delta = \left(\frac{\sum_{k} \mu_{C,k}^{2} }{\sum_{j}\mu_{C,j}^{2} + \sigma_{j}^{2}} - \frac{\sum_{k} \mu_{B,k}^{2} }{\sum_{j}\mu_{B,j}^{2} + \sigma_{j}^{2}} - Q(\alpha)(\frac{\sum_{k} \sigma_{k}^{4} + 2\mu_{B,k}^{2}\sigma_{k}^{2}}{(\sum_{j}\mu_{B,j}^{2} + \sigma_{j}^{2})^{2}})\right)\left(\frac{(\sum_{j} \mu_{C,j}^{2} + \sigma_{j}^{2})^{2}}{\sum_{k} \sigma_{k}^{4} + 2\mu_{C,k}^{2}\sigma_{k}^{2}}\right)^{1/2}
\end{equation}

While this expression is cumbersome, the key point is that the statistical power to discriminate classes of data points as a function of the parameters of the data (the mean and covariance) is computable. This formulation can be used to guide transformations of $\Sigma$ - representations such as metric learning that can embed the data into a new space that maximizes desirable statistical properties. It is worth noting that linear rescaling of the data to manipulate $\Sigma$ will also affect the means $\mathbf{\mu}$. This can be rectified by scaling the means by the variances $\sigma_{i}^{2}$ to make them dimensionless. Let the dimensionless means $\zeta, \eta$ be defined as $\mu_{C,i} = \zeta_{i}\sigma_{i}$, $\mu_{B,i} = \eta_{i}\sigma_{i}$. We can rewrite $\Delta$ purely in terms of the variances, treating the dimensionless means as parameters:
\begin{equation}
    \Delta = \left(\frac{\sum_{k} \sigma_{k}^{2}\zeta_{k}^{2}}{\sum_{j}\sigma_{j}^{2}(1+\zeta_{j}^{2})} - \frac{\sum_{k} \sigma_{k}^{2}\eta_{k}^{2}}{\sum_{j}\sigma_{j}^{2}(1 + \eta_{j}^{2})} - Q(\alpha)\left(\frac{\sum_{k} \sigma_{k}^{4}(1 + 2 \eta_{k}^{2})}{(\sum_{j}\sigma_{j}^{2}(1 + \eta_{j}^{2}))^{2}}\right)\right) \left(\frac{(\sum_{j} \sigma_{j}^{2}(1 + 2 \zeta_{j}^{2}))^{2}}{\sum_{k} \sigma_{k}^{4}(1 + 2\zeta_{k}^{2})}\right)^{1/2}
\end{equation}

The expression for $\Delta$, the quantile of $\cos(X, Y)$ at a particular power $\alpha$ assumes that $\cos(X, Y)$ has a normal distribution. For sufficiently large N, and with the assumptions needed for the approximation of the moments of cosine, this isn't unreasonable. However, similar discriminative statistics can be devised that rely just on the moments of the approximation, like the signal-to-noise ratio. This characterization of the distribution of $\cos(X, Y)$ enables optimization of the data embedding with respect to statistical power. 

\section*{Experiments}

\begin{figure}[ht]
\centering
\includegraphics[width=0.8\textwidth]{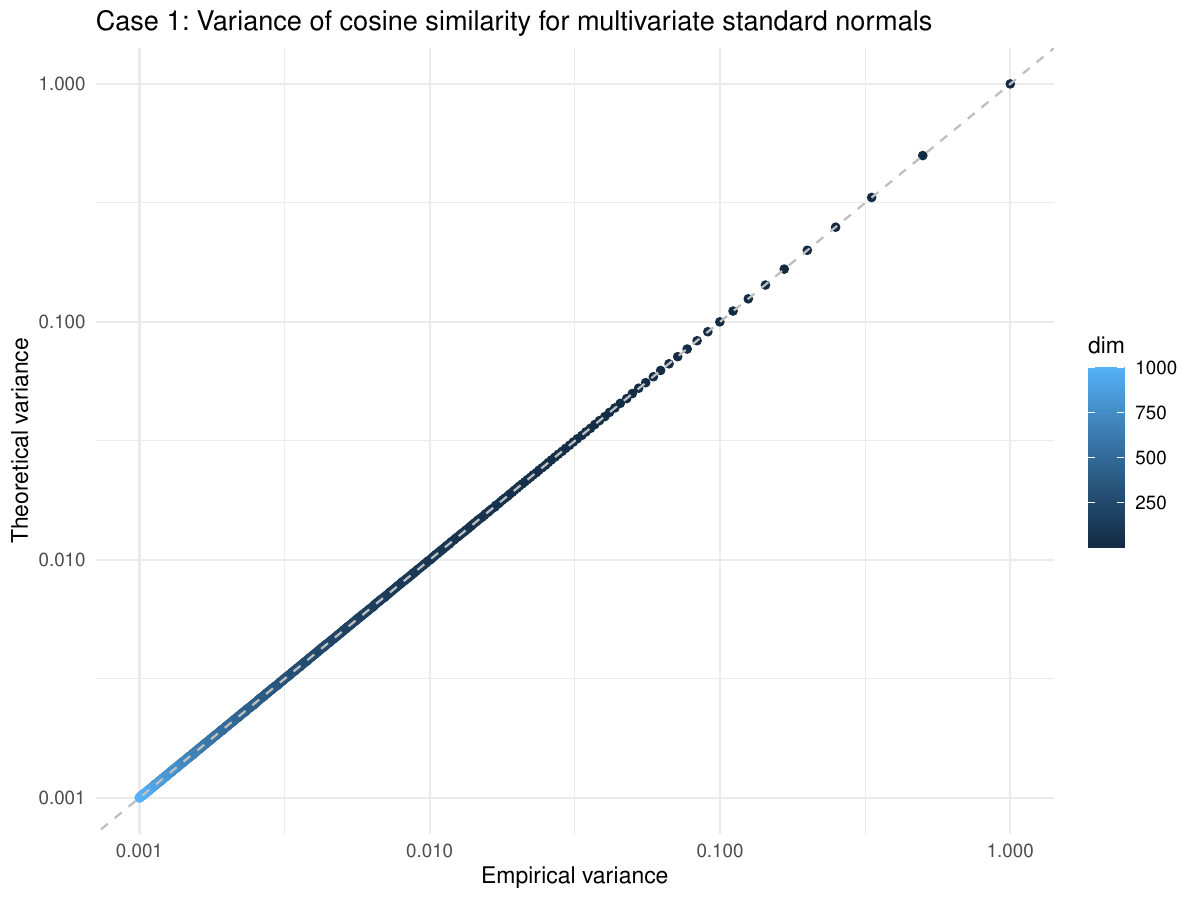}
\caption{Case 1: Comparison of variance of cosine similarity of multivariate standard normals with theory. As expected, experiment agrees perfectly with the theoretical variance. Cosine similarity is evaluated on N=1000 samples of standard multivariate normals (mean 0, variance 1) of dimension dim ranging from 1 to 1000. The variance follows 1/n, as in \ref{eqn:Case1}}
\label{Case1CosVar}
\end{figure}

To check the validity of these relationships and approximations, we next performed a series of simulations. For the first case, where $X \sim \mathcal{N}(\boldsymbol{0}, \mathbbm{1})$, for a particular value of N (the dimension of $X$), we sample 1000 vectors of length n where each component is a standard normal. We then compute the variance of cosine similarity between all pairs of these 1000 vectors and compare this variance to the predicted value from Equation \ref{eqn:Case1}. The simulation was evaluated at dimensions from 1 to 1000, and the empirical variance agrees with the prediction (Pearson = 0.9999997) (Figure \ref{Case1CosVar}).

\begin{figure}
    \begin{subfigure}{0.48 \textwidth}
    \includegraphics[width=\textwidth]{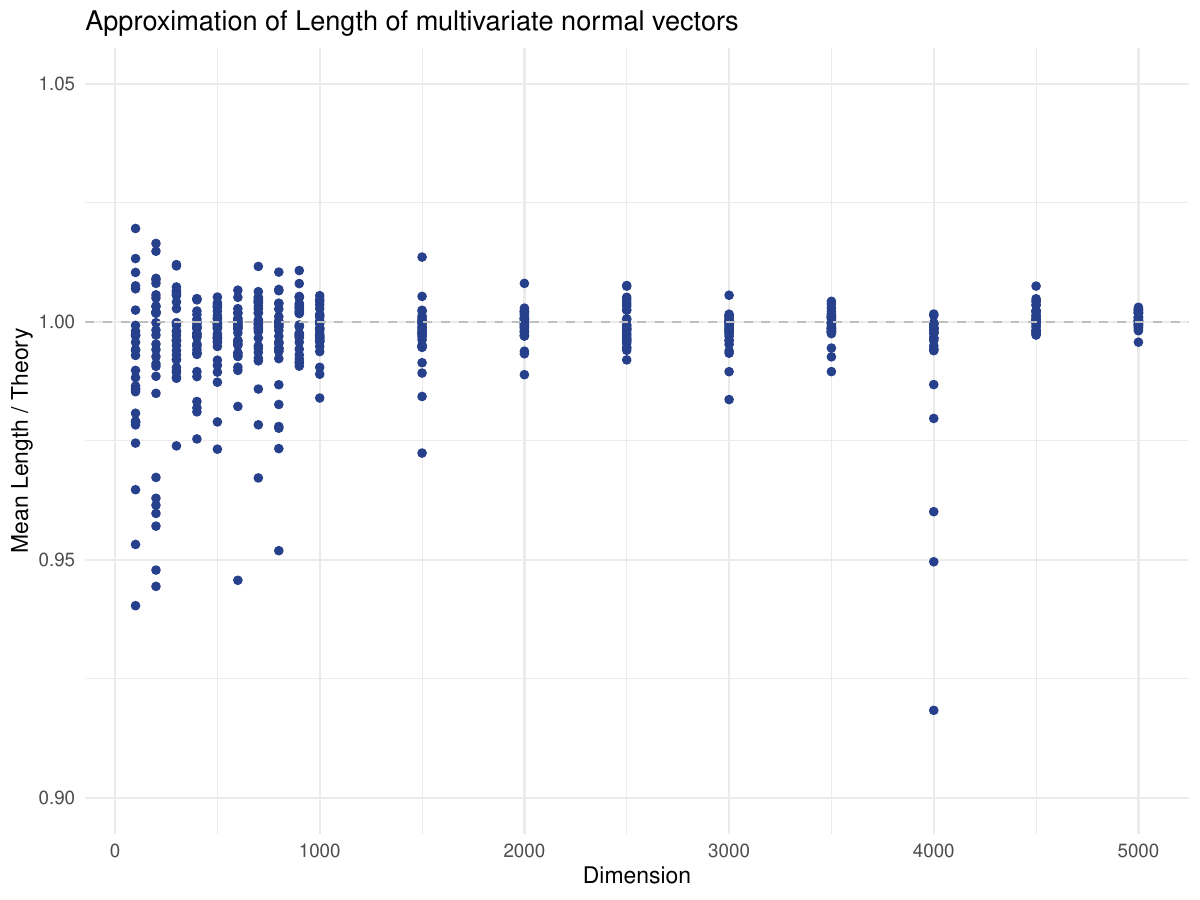}
    \caption{}
    \label{Case2NormMean}
    \end{subfigure}
    \begin{subfigure}{0.48 \textwidth}
    \includegraphics[width=\textwidth]{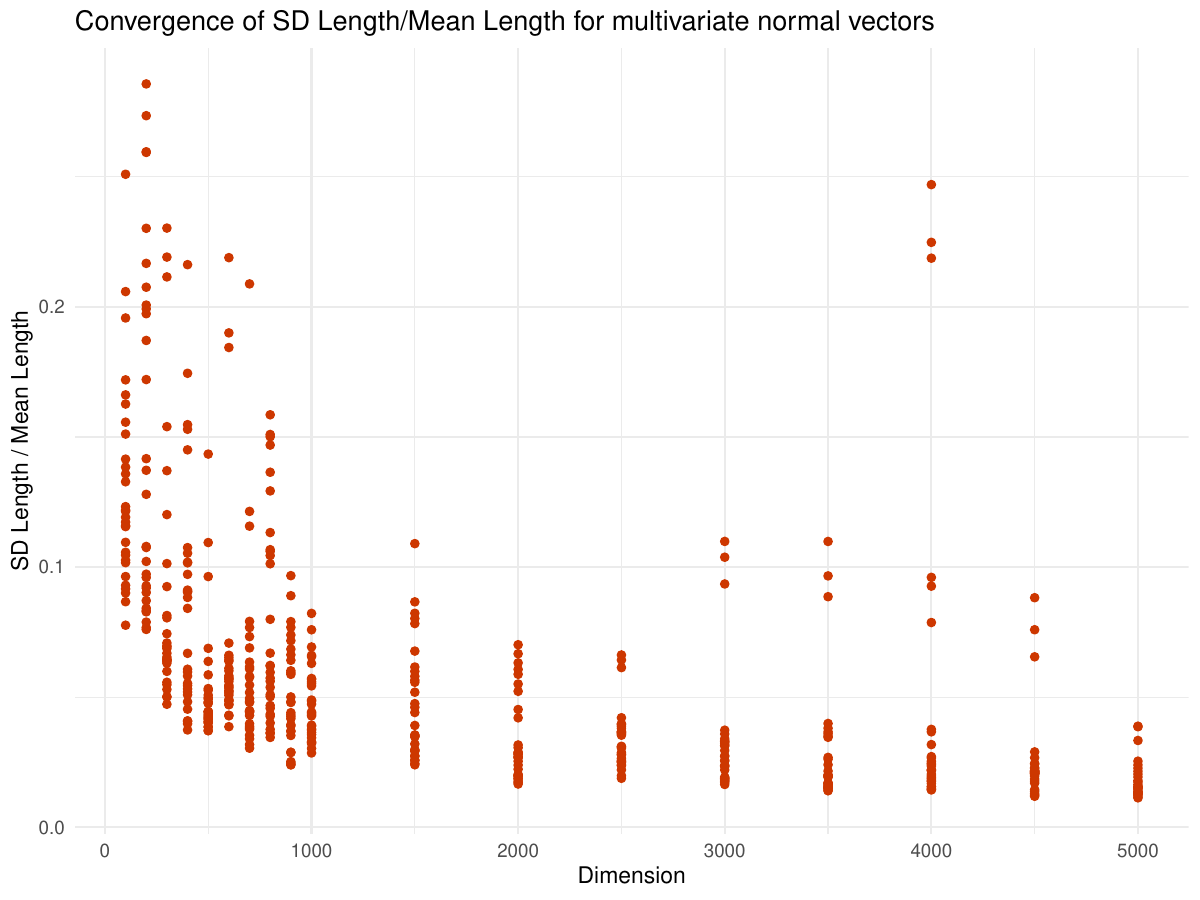}
    \caption{}
    \label{Case2NormSD}
    \end{subfigure}
    \caption{Approximation of L2-norm of centered multivariate normal. The approximation of Case 2 relies on approximating the norm of a vector as in Equation \ref{normapprox}. This simulation shows that for sufficiently large dimension, the approximation is good. (a) The ratio of the observed mean norm to the approximation tends to 1 as the dimension goes to infinity. (b) The variance of the observed norm divided by the observed mean tends to 0.}
\end{figure}

\begin{figure}
    \centering
    \includegraphics[width=0.8\textwidth]{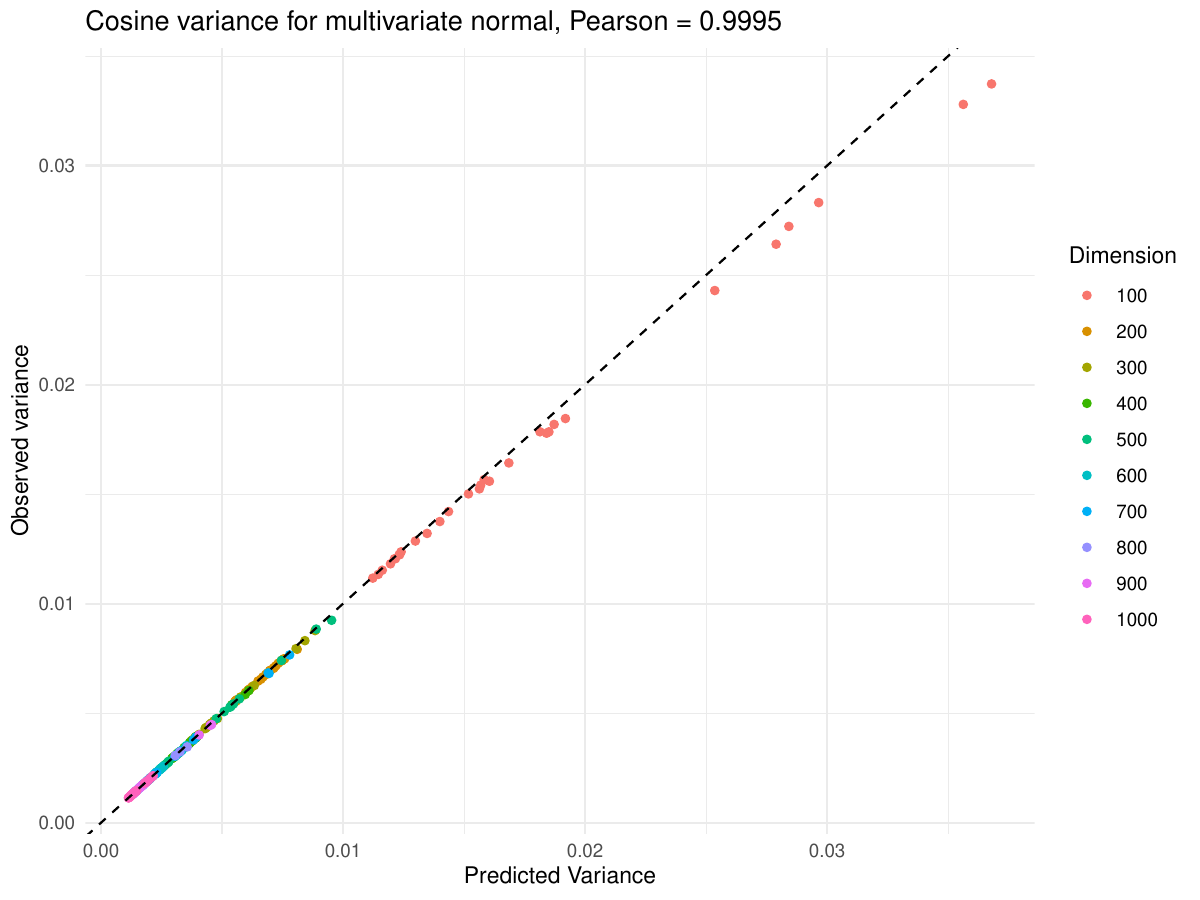}
    \caption{Case 2: Comparison of variance of cosine similarity of generalized centered multivariate normals with theory. In the tested range, where dimension varies from 100 to 1000, and with the gamma-distributed eigenvalues of the covariance matrix, theory agrees well with simulation, with a Pearson's correlation of 0.9995.}
    \label{Case2CosVar}
\end{figure}

\begin{figure}
    \centering
    \includegraphics[width=0.8\textwidth]{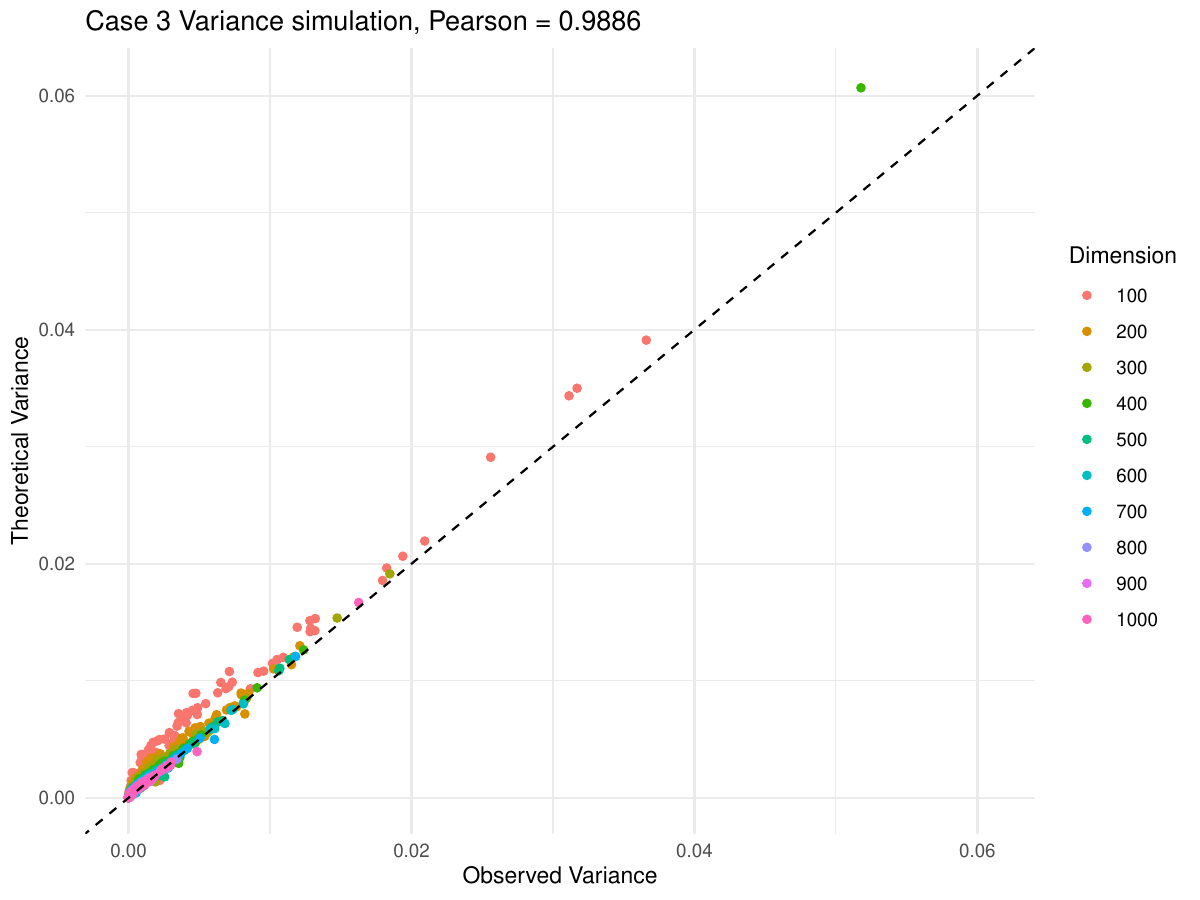}
    \caption{Case 3: Comparison of variance of cosine similarity of generalized multivariate normals with theory. The dimension varies from 100 to 1000 with gamma-distributed eigenvalues of the covariance matrix, normally distributed means, and 2000 sampled vectors. Theory agrees with simulation well, though with some deviations for small variance.}
    \label{Case3CosVar}
\end{figure}

Our approximation for Case 2, the centered multivariate distribution, relies on the approximation of the L2 norm or Euclidean length of a vector. To check this, we sampled multivariate normal distributions and computed the length of the vectors. First, we sampled two hyperparameters - the shape and rate - from a gamma distribution with shape = 1, rate = 2. The resulting distribution is exponential and was chosen to have a large range of values. We then sampled eigenvalues from a gamma distribution with shape and rate given by the hyperparameters. It is necessary to vary the distribution from which eigenvalues are sampled, or otherwise for sufficiently large n, the samples will have nearly identical eigenvalue distributions. We then sampled 100 vectors from a multivariate normal with covariance matrix with the given eigenvalues and computed the mean length and variance of the length. We repeated the entire procedure 20 times for each value of the dimension, and five times for each particular set of eigenvalues. 

The ratio of the empirical mean length to the approximation of the mean length, which was generated by a bound from Jensen's inequality (Appendix \ref{apdx:normMultinorm}), tends to 1 as the dimension gets sufficiently large (Figure \ref{Case2NormMean}). The empirical does sometimes exceed the theoretical upper bound, which we attribute to sampling error. The occasional distribution outlier also does occur, as shown in the simulation for dimension = 4000. The ratio of the standard deviation of the length to the mean length also tends to 0 as the dimension gets sufficiently large (Figure \ref{Case2NormSD}). This suggests that our approximation is reasonable, and in practice quite good for the regime of biological datasets. 

Next, we sought to empirically check the approximation of the variance of cosine similarity for Case 2. As with approximating the norm of the vectors, to allow variation in the eigenvalue distribution, we first sampled shape and rate hyperparameters from a gamma distribution with shape = 1, rate = 2. We then sampled n eigenvalues from a gamma distribution with shape and rate from the hyperparameters. We sampled 2000 multivariate normals with mean 0 and covariance matrix eigenvalues as sampled and computed the mean and variance of cosine similarity over the sample. We repeated this ten times for each eigenvalue sample and sampled ten eigenvalue hyperparameters for each value of dimension from 100 to 1000 at intervals of 100. In this simulation, the predicted variance and empirical variance had a Pearson's correlation of 0.9995, with the most deviation for N = 100 dimensions (Figure \ref{Case2CosVar}). This result suggests that for the practical regimes, the theoretical variance agrees well with experiment. 

Finally, we sought to empirically check the general approximation of the variance of cosine of Case 3, any multivariate distribution with finite mean and variance. As before, we sampled shape and rate hyperparameters from a gamma distribution with shape = 1, rate = 2. We then sampled N eigenvalues from a gamma distribution defined by the hyperparameters. We sampled 2000 multivariate normals with means drawn from $N(0,2)$ and covariance matrix eigenvalues as sampled and computed the mean and variance of cosine similarity over the sample. We sampled one hundred different distributions for each value of the dimension from 100 to 1000 at intervals of 100. In this simulation, the theoretical variance and empirical variance had a Pearson's correlation of 0.9886 (Figure \ref{Case3CosVar}). We speculate that some of the deviation can be explained by uncertainty in the mean and variance from the sample. Still, theory generally agrees with experiment. 

\section*{Discussion}

In this work, we derive the mean and variance of cosine similarity over data with finite mean and variance. For illustrative purposes, we also include explicit calculations for normal distributions, which while idealized, are a reasonable first approximation for datasets. Additionally, the calculations for the normal distribution include full characteristic functions, enabling evaluation of other aspects of the distribution besides the mean and variance. Concerning application of this analysis, for any particular data set, computing the variance of cosine similarity over that dataset is trivial. The benefit of this work is understanding how properties of the data space - in particular the eigenvalues of the covariance matrix - affect the similarity distribution. 

Cosine similarity is a commonly used tool for identifying similar states or perturbations in the biological domain, as a loss function for machine learning models, and as a similarity function for unsupervised clustering over scale-invariant data. Tools like representation learning and the advent of self-supervised learning rely on learning transformations or representations of the data that are more useful for downstream tasks like prediction and clustering. This work provides a theoretical basis for understanding how linear transformations can be useful. This is relevant for more complex non-linear models, such as those for image processing, because the analysis can be applied directly to the last layer of the embedding. Biological analyses that use cosine similarity rely on discriminating pairs of related data elements from the background of pairs, and in most practical situations, the signal-to-noise ratio is quite low. Insights that can guide representations that maximize statistical power are of substantial practical value to the community. 

Our result on conditions minimizing the variance of cosine similarity - isotropy (or equal variance) for centered data, and near-isotropy for general data - are a key finding of this work. Because the variance of cosine similarity over the background data is inversely related to statistical power, this suggests that power to identify related data points is \textbf{maximized} when all the relevant signals are put on the same scale - which is atypical on biological data. Furthermore, the extrapolation to power has caveats: adding features that are pure noise reduces variance but does not increase power. Nevertheless, the isotropic conjecture formalizes the concept that equalizing all the relevant axes of variation - or biological processes or programs for biological data - maximizes the discriminative power of a dataset.

Cosine similarity is a simple scale-invariant similarity function that is closely related to Pearson's correlation and is an intuitive function from linear algebra and introductory physics. It is then surprising that computing the distribution of cosine similarity for even simple probability distributions is so challenging. We hope this work will be a contribution to the understanding of data representations and similarity functions. 

\section*{Derivations}

\subsection{Properties of cosine similarity}

\subsubsection*{Invariance under orthogonal transformation}\label{cosProps}

Cosine is invariant under orthogonal matrix transformation of the space. Let $U$ be an orthogonal matrix, i.e. $UU^{T} = \mathbb{I}_{n}$. Then for $A' = UA$:
\begin{align*}
    \cos(A', B') &= cos(UA, UB) \\
            &= \frac{(UA)^{T}(UB)}{\sqrt{(UA)^{T}(UA)(UB)^{T}(UB)}} \\
            &= \frac{A^{T}U^{T}UB}{\sqrt{(A^{T}U^{T}UA)(B^{T}U^{T}UB}} \\  
            &= \frac{A^{T}B}{\sqrt{A^{T}AB^{T}B}} \\
            &= \cos(A,B)
\end{align*}
By the spectral theorem, we can rotate X into a basis where the covariance matrix is diagonal without loss of generality. 


\subsubsection*{Scale invariance}
It follows trivially that cosine is invariant under positive scalar multiplication. For $k > 0$:
\begin{equation*}
cos(kA,B) = \frac{kA^{T}B}{\abs{kA}\abs{B}} = cos(A, B) = cos(A,kB)
\end{equation*}

\subsection{Relationship between cosine similarity and Pearson's correlation}

This work specifically addresses cosine similarity, defined as in Equation \ref{eqn:cosine}. Pearson's correlation $\rho$ is closely related:
\begin{equation}
    \rho(X,Y) = \frac{\mathbb{E}[(X - \mu_{X})(Y - \mu_{Y})]}{\sqrt{\mathbb{E}[X^{2}] - \mathbb{E}[X]^{2}}\sqrt{\mathbb{E}[Y^{2}] - \mathbb{E}[Y]^{2}}}
\end{equation}
When $\mathbb{E}[X] = \mathbb{E}[Y] = 0$, Pearson's correlation equals cosine similarity:
\begin{equation}
    \rho\left(X,Y | \mathbb{E}[X],\mathbb{E}[Y] = 0\right) = \frac{\mathbb{E}[XY]}{\sqrt{\mathbb{E}[X^{2}]\mathbb{E}[Y^{2}]}} = \frac{(1/N) X^{T}Y}{(1/N) \abs{X}\abs{Y}} = \frac{X^{T}Y}{\abs{X}\abs{Y}} = \cos(X,Y)
\end{equation}

Pearson's correlation then is just cosine similarity when the data is projected onto the (n-1) dimensional subspace where $\sum_{i} X_{i}$ = 0. Hence, the findings presented here for cosine similarity on an n-dimensional space apply for Pearson's correlation on an (n+1) dimensional space. As Pearson's correlation is perhaps the most commonly used correlation coefficient, this expands the scope of this work. 

\subsection{Derivation of Case 1: multivariate standard normal}\label{case1Deriv}

Let $X \sim \mathcal{N}(\boldsymbol{0}, \mathbbm{1})$, with an identity covariance matrix and dimension $N$. As shown above, cosine is scale invariant, so for $A, B \sim X$ i.i.d.:
\begin{equation}
    \cos(A,B) = \cos(\frac{A}{|A|}, \frac{B}{|B|})
\end{equation}
Therefore, we can consider the distribution of cosine similarity over $\frac{X}{|X|}$, which by Muller and Marsaglia is a uniform probability distribution over the unit sphere, denoted as $U(S^{n-1})$ \cite{muller_note_1959}. 

In the special case of the uniform distribution on $S^{n-1}$, we can exploit the spherical symmetry of the probability distribution when computing the distribution of $\cos(A,B)$. Because of spherical symmetry, one of the elements $B$ can be fixed WLOG, and the cosine similarity then equals the marginal distribution of $U(S^{n-1})$ along any axis. Fix $B$ along the first axis, i.e. $B = (1, 0, ... 0)$. Then, where $A = (X_{1}, X_{2}, ... X_{n})$, and noting that $\abs{A} = \abs{B} = 1$:
\begin{equation}
\cos(A,B) = A_{1} 
\end{equation}
The probability distribution of $X_{1}$ is the marginal distribution of $U(S^{n-1})$ along the first axis. From the definition of $X$, where $Z_{i} \sim \mathcal{N}(0,1)$, we have:
\begin{equation}
    X_{1} = \frac{Z_{1}}{\sqrt{\sum_{i=1}^{n} Z_{i}^{2}}}
\end{equation}
It is useful to consider the distribution of $X_{1}^{2}$:
\begin{equation}
    X_{1}^{2} = \frac{Z_{1}^{2}}{\sum_{i=1}^{n} Z_{i}^{2}} = \frac{Z_{1}^{2}}{Z_{1}^{2} + \sum_{i=2}^{n} Z_{i}^{2}}
\end{equation}
$Z_{1}^{2} \sim \chi^{2}(1)$, and $\sum_{i=1}^{n} Z_{i}^{2} \sim \chi^{2}(n-1)$. This function of independent $\chi^{2}$ random variables has a beta distribution of $X_{1}^{2} \sim Beta(\frac{1}{2}, \frac{n-1}{2})$\cite{soch_relationship_2022}. 
Then:
\begin{align}
f_{X_{1}}(x) &= f_{X_{1}^{2}}(x^{2}) \abs{\frac{d}{dx} x^{2}} \\
             &= \frac{1}{B(\frac{1}{2}, \frac{n-1}{2})} (x^{2})^{-1/2} (1-x^2)^{(n-3)/2} 2x \\
             &= \frac{2}{B(\frac{1}{2}, \frac{n-1}{2})} (1-x^{2})^{(n-3)/2}
\end{align}
Making the substitution $u = (1+x)/2$, $x^{2} = (2u - 1)^2$ (note the extra 2 in the numerator comes from the Jacobian of the transform):
\begin{align}
f_{U}(u) &= \frac{2}{B(\frac{1}{2}, \frac{n-1}{2})} (1 - (2u-1)^2)^{(n-3)/2} 2 \\
         &= \frac{1}{B(\frac{1}{2}, \frac{n-1}{2})} 2^{n-2} (u-u^{2})^{(n-3)/2} \\
         &= \frac{1}{B(\frac{1}{2}, \frac{n-1}{2})} 2^{n-2} u^{(n-1)/2 - 1}(1-u)^{(n-1)/2 - 1} \\
         &\sim Beta(\frac{N-1}{2}, \frac{n-1}{2})
\end{align}
$U = (1+X_{1})/2$ is a Beta distributed random variable with parameters $\alpha=\frac{N-1}{2}, \beta=\frac{n-1}{2}$. The variance of $X_{1}$, from the properties of a beta distribution is:
\begin{equation}
    \sigma_{X_{1}}^{2} = 2^{2} \frac{\alpha\beta}{(\alpha+\beta)^{2}(\alpha + \beta + 1)} = \frac{(n-1)^{2}}{n(n-1)^2} = \frac{1}{n}
\end{equation}

\subsection{Derivation of Case 2: centered multivariate distribution}\label{case2Deriv}

\subsubsection{The multivariate normal case}
Case 2: a more general problem is a centered multivariate distribution with finite variances that can be unequal. Without loss of generality, let $\Sigma$ be diagonal with entries $\Sigma_{ii} = \sigma_{i}^{2}$. First, we consider the special case where $X$ is a multivariate normal:
\begin{align}
X &\sim \mathcal{N}(0, \Sigma) \\
X_{i} &\sim \mathcal{N}(0, \sigma_{i}^{2})
\end{align}
Consider the cosine of two such variables $X,Y \sim P_{X}$ i.i.d. Each term in the numerator is a product of two independent normally distributed random variables, and can be written as a function of the sums of squares of normals. 
\begin{align}
X_{i} + Y_{i} &\sim \mathcal{N}(0, 2\sigma_{i}^{2})\\
(X_{i} + Y_{i})^2 &\sim \Gamma(\frac{1}{2}, \frac{1}{4 \sigma_{i}^{2}}) \\
X_{i}Y_{i} &= \frac{1}{4}[(X_{i} + Y_{i})^{2} - (X_{i} - Y_{i})^{2}] 
\end{align}
This probability distribution of the product of two normally distributed random variables turns out to be a Bessel function of the second kind\cite{wishart_distribution_1932}. The variance of the numerator of cosine similarity follows in a straightforward way as the sum of independent random variables.
\begin{align*}
Var(X_{i}Y_{i}) &= \mathbb{E}[(X_{i}Y_{i})^{2}] - \mathbb{E}[X_{i}Y_{i}]^2 \\
                &= \mathbb{E}[X_{i}^{2}]\mathbb{E}[Y_{i}^{2}] - \mathbb{E}[X_{i}]^{2}\mathbb{E}[Y_{i}]^{2} \\
                &= Var(X_{i})Var(Y_{i}) + \mathbb{E}[Y_{i}]^{2}Var(X_{i}) + \mathbb{E}[X_{i}]^{2} Var(Y_{i}) \\
                &= Var(X_{i})Var(Y_{i}) = \sigma_{i}^4
\end{align*}
\begin{equation}
    Var(\sum_{i = 1}^{k} X_{i} Y_{i}) = \sum_{i} \sigma_{i}^4
\end{equation}
However, the denominator of cosine makes computing the variance of the total expression challenging, as it is the product of the norms of the vectors. The argument under the square root of the norm for multivariate Gaussian vectors with i.i.d. standard normal components is a chi-squared random variable, but once the variances $\sigma_{i}^{2}$ of each component are allowed to vary, the argument of the square root becomes a linear combination of chi-squared random variables, and a closed analytic expression for the probability density function is not known\cite{bausch_efficient_2013}. This is a special case of the generalized chi-squared distribution. 

The key simplification is to relax the normalization implicit in the cosine similarity function and instead rescale the input vectors so the \textbf{expectation} of the squared norm is 1, rather than guaranteeing the norm to be 1. Equivalently, we can replace the norms of the random variables $X,Y$ in the denominator of the expression for cosine with the roots of the expectation value of the squared norms. Let $X \sim \mathcal{N}(0, \Sigma_{X})$, where $\Sigma_{X}$ is diagonal with $\Sigma_{X,ii} = \sigma_{i}^{2}$. 
\begin{align}
    b(X) &\equiv \frac{X}{\sqrt{\sum_{i} \sigma_{i}^2}}  \\
    \mathbb{E}[\abs{b(X)}^{2}] &= \frac{\sum_{i} \sigma_{i}^{2}}{\sum_{i}\sigma_{i}^{2}} = 1 \\
    \Sigma_{b, ij} &\equiv \lambda_{i} = \mathbb{E}[\frac{1}{\sum \sigma_{i}^{2}} (X_{i} - \overline{X_{i}})(X_{j} - \overline{X_{j}})] = \frac{1}{\sum \sigma_{i}^{2}} \Sigma_{X, ij}
\end{align}
A note on the second line, it is straightforward to calculate the expectation of the squared norm $\abs{X}^{2}$ (See Derivations \ref{normMultivar}), but trickier to calculate the expectation of the norm. We can then approximate cosine with $\widehat{\cos}$ defined as:
\begin{equation}
    \cos(X,Y) = \frac{X^{T}Y}{\abs{X}\abs{Y}} \approx \frac{X^{T}Y}{\sqrt{E[\abs{X}]^{2}E[\abs{Y}]^{2}}} = b(X)^{T} b(Y) \equiv \widehat{\cos}(X,Y)
\end{equation}
We can compute the moments of $H = \widehat{\cos}(X,Y)$. Let the nth moment be $<H^{n}> = \mathbb{E}[H^{n}]$. Using Equation \ref{eqn:charProd}:
\begin{equation}
    \phi_{H} = \prod_{i} (1 + \lambda_{i}^{2} t^{2})^{-\frac{1}{2}}
\end{equation}
\begin{align}
    <H> &= i^{-1} \frac{d}{dt} \phi_{H} |_{t = 0} \\
       &= i^{-1} \sum_{i} \prod_{j \neq i} (1 + \lambda_{j}^{2} t^{2})^{-\frac{1}{2}} \frac{d}{dt} [(1 + \lambda_{i}^{2}t^{2})^{-\frac{1}{2}}] \\
       &= i^{-1} \sum_{i} \prod_{j \neq i} (1 + \lambda_{j}^{2} t^{2})^{-\frac{1}{2}} (-\lambda_{i}^{2}t)(1 + \lambda_{i}^{2}t^{2})^{-\frac{3}{2}} \\
       &= 0
\end{align}
For conciseness, let $A_{j} = (1 + \lambda_{j}^{2} t^{2})^{-\frac{1}{2}}$. 
\begin{align}
    <H^{2}> &= i^{-2} \frac{d}{dt} \left( \sum_{i} \prod_{j \neq i} (1 + \lambda_{j}^{2} t^{2})^{-\frac{1}{2}} (-\lambda_{i}^{2}t)(1 + \lambda_{i}^{2}t^{2})^{-\frac{3}{2}} \right) |_{t = 0} \\ 
    &= - \sum_{i} \prod_{j \neq i} A_{j} \frac{d}{dt} \left( (-\lambda_{i}^{2}t)(1 + \lambda_{i}^{2}t^{2})^{-\frac{3}{2}} \right) + 2\sum_{i,j\neq i} \prod_{k \neq i,j} A_{k} (1 + \lambda_{i}^{2}t^{2})^{-\frac{3}{2}}(1 + \lambda_{j}^{2}t^{2})^{-\frac{3}{2}}\lambda_{i}^{2}\lambda_{j}^{2}t^{2} \\
    &= - \sum_{i} \prod_{j \neq i} A_{j} [-\lambda_{i}^{2}(1+\lambda_{i}^{2}t^{2})^{-\frac{3}{2}} + \lambda_{i}^{2} t(\frac{3}{2})(1 + \lambda_{i}^{2}t^{2})^{-\frac{5}{2}}2 \lambda_{i}^{2}t + 2\sum_{i,j\neq i} \prod_{k \neq i,j} A_{k} (1 + \lambda_{i}^{2}t^{2})^{-\frac{3}{2}}(1 + \lambda_{j}^{2}t^{2})^{-\frac{3}{2}}\lambda_{i}^{2}\lambda_{j}^{2}t^{2} \\
    &= \sum_{i} \lambda_{i}^{2}
\end{align}
Plugging in the definition of $\lambda_{i}$ as the normalized variances:
\begin{align}
    Var(H) &= <H^{2}> - <H>^{2} = \sum_{i} \lambda_{i}^{2} \\
    Var(H) &= \sum_{i} \frac{\sigma_{i}^{4}}{(\sum_{j} \sigma_{j}^{2})^{2}}
\end{align}
For convenience, let $v_{i} = \sigma_{i}^{2}$. To minimize $Var(H)$, compute the partial derivative with respect to $v_{i}$, as $Var(H)$ is strictly convex per \eqref{eq:FormulaPhiHatKN}:
\begin{align}
    Var(H) &= \sum_{i} \frac{v_{i}^{2}}{(\sum_{j} v_{j})^{2}} \\ 
    \frac{\partial Var(H)}{\partial v_{i}} &= 2 (\sum_{j} v_{j})^{-2} v_{i} - 2 (\sum_{j} v_{j})^{-3} \left(\sum_{j} v_{j}^{2}\right) = (\sum_{j} v_{j}) v_{i} - (\sum_{j} v_{j}^{2})
\end{align}
The value of $v_{i}$ that minimizes $Var(H)$ yields:
\begin{equation}
    \tilde{v_{i}} = \frac{\sum_{j \neq i} v_{i}^{2}}{\sum_{j \neq i} v_{i}}
\end{equation}
This shows that the global minimum for $Var(H)$, the approximation of $Var(\cos(X,Y))$, is achieved when $\sigma_{i}^{2} = \sigma_{j}^{2}, \forall i, j$.  
\subsubsection{Case 2 without assuming normality}
Generalizing beyond normality: for any probability distribution of $X$ with mean $0$ and finite covariance $\Sigma$, this rescaling trick is equivalent to the approximation of the squared norm of the arguments with the expected value of the squared norm. For clarity, we denote this approximation as $\widehat{\cos}$. This is explored at greater length in Derivations \ref{normMultivar}. 
\begin{equation}
    \cos(X,Y) = \frac{X^{T}Y}{\abs{X}\abs{Y}} \approx \frac{X^{T}Y}{\sqrt{\mathbb{E}[\abs{X}^{2}]\mathbb{E}[\abs{Y}^{2}]}} \equiv \widehat{\cos}(X,Y)
\end{equation}
Without specifying the distribution $P_{X}(X)$, it is not possible to write down the characteristic function of $\cos(X,Y)$, for $X,Y \sim P_{X}(X)$ i.i.d. However, it is possible to compute the first moments of $\cos(X,Y)$ using the approximation. Note that the last line follows because $\mathbb{E}[X_{k}] = 0$. 
\begin{align}
    \mathbb{E}[\widehat{\cos}(X,Y)] &= \frac{1}{\sqrt{(\sum_{k}\sigma_{k}^{2})^{2}}} \mathbb{E}[\sum_{k} X_{k}Y_{k}] \\
                                    &= \frac{1}{(\sum_{k}\sigma_{k}^{2})} \sum_{k}\mathbb{E}[X_{k}Y_{k}] \\
                                    &= \frac{1}{(\sum_{k}\sigma_{k}^{2})} \sum_{k}\mathbb{E}[X_{k}]\mathbb{E}[Y_{k}] \\
                                    &= 0
\end{align}
\begin{align}
    \mathbb{E}[\widehat{\cos}(X,Y)^{2}] &= \frac{1}{(\sum_{i}\sigma_{i}^{2})^{2}} \mathbb{E}[(X^{T}Y)^{2}] \\
        &= \frac{1}{(\sum_{i}\sigma_{i}^{2})^{2}} (\mathbb{E}[\sum_{k} X_{k}^{2}Y_{k}^{2}] + \mathbb{E}[\sum_{k}\sum_{j\neq k} X_{k}Y_{k}X_{j}Y_{j}]) \\
        &= \frac{1}{(\sum_{i}\sigma_{i}^{2})^{2}} (\sum_{k} \mathbb{E}[X_{k}^{2}]\mathbb{E}[Y_{k}^{2}] + \sum_{k}\sum_{j \neq k} \mathbb{E}[X_{k}Y_{k}X_{j}Y_{j}]) \\
        &= \frac{1}{(\sum_{i}\sigma_{i}^{2})^{2}} (\sum_{k} \sigma_{k}^{4})
\end{align}
\begin{equation}
    Var(\widehat{\cos}(X,Y)) = \sum_{k} \frac{\sigma_{k}^{4}}{(\sum_{j} \sigma_{j}^{2})^{2}}
\end{equation}
This is the same result as with centered multivariate normals, showing that the solution generalizes to any multivariate distribution with zero mean and finite variance. As before, the minimum value for the variance of $\widehat{\cos}(X,Y)$ is achieved when $\sigma_{i}^{2} = \sigma_{j}^{2} \forall i, j$.

\subsection{Derivation of Case 3: generalized multivariate distribution}\label{case3Deriv}
Case 3: The most general case for a multivariate distribution removes the requirement that the mean be 0 and does not require the eigenvalues to be equal. 
\subsubsection{Multivariate normal}
First, consider the multivariate normal case. WLOG, let $\Sigma$ be diagonal with entries $\Sigma_{ii} = \sigma_{i}^{2}$, and let the mean of $X_{i}$ be $\mu_{i}$.
\begin{align}
    X &\sim N(\mu, \Sigma) \\
    X &\sim N(\mu_{i}, \sigma_{i}^{2})
\end{align}
We extend the approximation made in case 2. For $A, B \sim X$:
\begin{equation}
    \cos(A,B) = \frac{\sum_{j} A_{j} B_{j}}{|A||B|}
\end{equation}

From Appendix Equations \ref{normLengthMean}, \ref{normLengthVar}, we have the following approximations of mean length and variance of $\abs{A}$:
\begin{align}
    \mathbb{E}[\abs{A}] &\leq \sqrt{\sum_{j} \mu_{j}^{2} + \sigma_{j}^{2}} \\
    \mathbb{E}[\abs{A}^{2}] &= \sum_{k} \sigma_{k}^{2} + \mu_{k}^{2}
\end{align}

Assuming the $\sigma_{i}^{2}$s are distributed such that Lindeberg's condition is satisfied (Derivations \ref{normMultivar}), or more generally that the contribution to the total variance of any one eigenvalue is arbitrarily small for sufficiently large values of n, we can employ the same trick as with case 2. The reasoning is then that for sufficiently large n and with constraints on the $\sigma_{i}^{2}$, the norm of a multivariate normal vector from $X$ is a constant. 
Using this approximation, the distribution of cosine can be simplified to ($\widehat{\cos}$ denoting the approximation):
\begin{equation}
\widehat{\cos}(A,B) = \frac{1}{\sum_{j}\mu_{j}^{2} + \sigma_{j}^{2}} \sum_{j}A_{j}B_{j}
\end{equation}
Let $C = \frac{1}{\sum_{j}(\mu_{j}^{2} + \sigma_{j}^{2})}$. The characteristic function $\phi_{\widehat{\cos}}$ for the multivariate normal, per Equation \ref{eqn:charProd}, is then:
\begin{equation}
    \phi_{\widehat{\cos}} = \prod_{k} (1 + \sigma_{k}^{4}C^{2}t^{2})^{-1/2} \exp\left(\frac{i\mu_{k}^{2}Ct}{(1 + \sigma_{k}^{4}C^{2}t^{2})} - \frac{\mu_{k}^{2}\sigma_{k}^{2}C^{2}t^{2}}{(1 + \sigma_{k}^{4}C^{2}t^{2})} \right) 
\end{equation}

The moments of $\widehat{\cos}$ for the generalized multivariate normal are then:
\begin{align*}
    <\widehat{\cos}> &= \frac{1}{i}\frac{d}{dt}\phi_{\widehat{\cos}}(t) \\
           &= \frac{1}{i}\sum_{k}\phi_{\widehat{\cos}}(t) \left[\frac{i\mu_{k}^{2}C - 2\mu_{k}^{2}\sigma_{k}^{2}C^{2}t^{2} - \sigma_{k}^{4}C^{2}t}{(1 + \sigma_{k}^{4}C^{2}t^{2})} + \frac{2(\mu_{k}^{2}\sigma_{k}^{6}C^{3}t^{3} - i\mu_{k}^{2}\sigma_{k}^{4}C^{3}t^{2})}{(1 + \sigma_{k}^{4}C^{2}t^{2})^{2}}\right] \\
    <\widehat{\cos}> &= \frac{\sum_{k}\mu_{k}^{2}}{\sum_{j}\mu_{j}^{2} + \sigma_{j}^{2}}
\end{align*}
\begin{align*}
    <\widehat{\cos}^{2}> &= (-1)\frac{d^{2}}{dt^{2}}\phi_{cos}(t) \\
           &= \frac{1}{(\sum_{j}\mu_{j}^{2} + \sigma_{j}^{2})^{2}}\left(\sum_{k} (\sigma_{k}^{4} + 2\mu_{k}^{2}\sigma_{k}^{2} + \mu_{k}^{4}) + \sum_{j}\sum_{k \neq j} \mu_{j}^{2}\mu_{k}^{2}\right)
\end{align*}

From this, the variance of the cosine similarity for the generalized multivariate normal is:
\begin{equation}
    Var(\widehat{\cos}) = \frac{\sum_{k} \sigma_{k}^{2} (\sigma_{k}^{2} + 2 \mu_{k}^{2})}{(\sum_{j} \mu_{j}^{2} + \sigma_{j}^{2})^{2}}
\end{equation}

\subsubsection{Generalized multivariate distribution without normality}
The reasoning can be generalized to any probability distribution of $X$ with finite mean $\mu$ and covariance $\Sigma$. While we cannot derive the characteristic function for an unknown probability distribution, we can calculate the moments with the same reasoning as before:
\begin{align}
    \mathbb{E}[\widehat{\cos}(X,Y)] &= \frac{1}{\sqrt{(\sum_{k}(\sigma_{k}^{2} + \mu_{k}^{2}))^{2}}} \mathbb{E}[\sum_{k} X_{k}Y_{k}] \\
        &= \frac{1}{(\sum_{k}(\sigma_{k}^{2} + \mu_{k}^{2}))} \sum_{k}\mathbb{E}[X_{k}Y_{k}] \\
        &= \frac{1}{(\sum_{k}(\sigma_{k}^{2} + \mu_{k}^{2}))} \sum_{k}\mathbb{E}[X_{k}]\mathbb{E}[Y_{k}] \\
        &= \frac{\sum_{k}\mu_{k}^{2}}{(\sum_{j}(\sigma_{j}^{2} + \mu_{j}^{2}))}
\end{align}
\begin{align}
    \mathbb{E}[\widehat{\cos}(X,Y)^{2}] &= \frac{1}{(\sum_{k}\sigma_{k}^{2} + \mu_{k}^{2})^{2}} \mathbb{E}[(X^{T}Y)^{2}] \\
        &= \frac{1}{(\sum_{k}\sigma_{k}^{2} + \mu_{k}^{2})^{2}} (\mathbb{E}[\sum_{k} X_{k}^{2}Y_{k}^{2}] + \mathbb{E}[\sum_{k}\sum_{j\neq k} X_{k}Y_{k}X_{j}Y_{j}]) \\
        &= \frac{1}{(\sum_{k}\sigma_{k}^{2} + \mu_{k}^{2})^{2}} (\sum_{k} \mathbb{E}[X_{k}^{2}]\mathbb{E}[Y_{k}^{2}] + \sum_{k}\sum_{j \neq k} \mathbb{E}[X_{k}Y_{k}X_{j}Y_{j}]) \\
        &= \frac{1}{(\sum_{k}\sigma_{k}^{2} + \mu_{k}^{2})^{2}} (\sum_{k} (\sigma_{k}^{2} + \mu_{k}^{2})^{2} + \sum_{k}\sum_{j\neq k} \mu_{k}^{2}\mu_{j}^{2})
\end{align}
As before, this is the same result as for the multivariate normal distribution, meaning that assuming finite mean, finite variance, and sufficiently high dimension with Lindeberg's condition for the approximation to hold:
\begin{equation}
    Var(\widehat{\cos}) = \frac{\sum_{k} \sigma_{k}^{2} (\sigma_{k}^{2} + 2 \mu_{k}^{2})}{(\sum_{j} \mu_{j}^{2} + \sigma_{j}^{2})^{2}}
\end{equation}

\subsubsection{Minimizing the variance of cosine of the generalized multivariate distribution}

Cosine is not invariant under general transformation of the data space, and it is reasonable to minimize the variance of $\widehat{\cos}$ with respect to the eigenvalues of the covariance matrix, $\sigma_{i}^{2}$. This is equivalent to transforming the data space with a positive definite matrix, or linearly embedding the data in some new space. Because such a transformation would also change the means, it is useful to represent $Var(\cos)$ with dimensionless means, $\mu_{i} = \eta_{i} \sigma_{i}$. With this change of variables, the $\eta$ values are constants, and the variance of $\cos$ depends only on the variance of the data. We exploit the convexity of the approximation of the variance as from \eqref{eq:FormulaPhiHatKN}. 
\begin{equation}
    Var(\widehat{\cos}) = \frac{\sum_{k} \sigma_{k}^{4} (1 + 2 \eta_{k}^{2})}{(\sum_{j} \sigma_{j}^{2} (1 + \eta_{j}^{2}))^{2}}
\end{equation}
\begin{equation}
    \frac{d Var(\widehat{\cos})}{d\sigma_{i}} = (-4)\frac{\sum_{k}\sigma_{k}^{4}(1 + 2 \eta_{k}^{2})}{\left(\sum_{j} \sigma_{j}^{2} (1 + \eta_{j}^{2}) \right)^{3}}\sigma_{i}(1 + \eta_{i}^{2})+ \frac{4\sigma_{i}^{3} (1 + 2 \eta_{i}^{2})}{\left(\sum_{j} \sigma_{j}^{2} (1 + \eta_{j}^{2}) \right)^{2}} 
\end{equation}
Setting the derivative equal to zero to find the minimum and simplifying gives:
\begin{align*}
    0 &= \left(\sum_{j} \sigma_{j}^{2} (1 + \eta_{j}^{2}) \right)\sigma_{i}^{2} (1 + 2\eta_{i}^{2}) - (1 + \eta_{i}^{2})\left( \sum_{k}\sigma_{k}^{4}(1 + 2 \eta_{k}^{2}) \right) \\ 
    &= \sum_{k} \sigma_{k}^{2}\left(\sigma_{i}^{2} (1 + \eta_{k}^{2})(1 + 2\eta_{k}^{2}) - \sigma_{k}^{2}(1 + 2\eta_{k}^{2})(1 + \eta_{i}^{2}) \right)
\end{align*}
The variance is minimized when, for all values of i and k:
\begin{equation}
\sigma_{i}^{2} \frac{1 + 2 \eta_{i}^{2}}{1 + \eta_{i}^{2}} = \sigma_{k}^{2} \frac{1 + 2 \eta_{k}^{2}}{1 + \eta_{k}^{2}}
\end{equation}
The $\sigma_{i}^{2}$ can be arbitrarily multiplied by a positive constant collectively because of the scale invariance of cosine. Defining $w_{i} = \frac{1 + \eta_{i}^{2}}{1 + 2\eta_{i}^{2}}$, and arbitrarily choosing a scale constant C, the minimum of the variance of cosine is achieved when the following is true:
\begin{equation}
    C \equiv \frac{\sigma_{i}^{2}}{w_{i}} =  \frac{\sigma_{j}^{2}}{w_{j}} \equiv \frac{1 + 2 \eta_{i}^{2}}{1 + \eta_{i}^{2}}\sigma_{i}^{2}
\end{equation}
\begin{equation}
    \sigma_{i}^{2} = C w_{i} = C \frac{1 + \eta_{i}^{2}}{1 + 2\eta_{i}^{2}}
\end{equation}
The minimum variance $\hat{Var}$ under rescaling given dimensionless means $\eta$ is:
\begin{equation}
    \hat{Var}(\widehat{\cos}) = \sum_{k} \frac{(1 + \eta_{k}^{2})^{2}}{1 + 2 \eta_{k}^{2}}\left(\sum_{j} \frac{(1 + \eta_{j}^{2})^{2}}{1 + 2\eta_{j}^{2}} \right)^{-2} = \frac{\sum_{k} w_{k}^{2}(1 + 2 \eta_{k}^{2})} {\left(\sum_{j}w_{j}(1 + \eta_{j}^{2}) \right)^{2}}
\end{equation}
The interpretation is interesting: $w_{i} \in [1/2, 1]$ is maximal when $\eta_{i} = 0$, and minimal as $\eta_{i} \rightarrow \infty$. But even with a dimension with arbitrarily large mean, the maximum variation between optimal eigenvalues is a factor of 2. Put another way, in the limit as $\eta_{i} \rightarrow \infty$, the variance $\sigma_{i}^{2} \rightarrow \frac12\sigma_{j}^{2}$ for $\eta_{j} = 0$. The variance along an axis with larger mean tends towards half that of an axis with zero mean. As a sanity check, when $\eta_{i} = 0 \forall i$, the minimization criterion becomes the isotropic $\sigma_{i}^{2} = \sigma_{j}^{2} \forall i, j$, and $\hat{Var} = \frac{1}{N}$. 

\subsection{Formalism of the approximation}

In this section, we establish the formalism behind the approximation of cosine used in the preceding sections. In the following we will consider the means as fixed values and let the variances vary over compact (closed and bounded) subsets of $[0,\infty)^N$. We will impose mild conditions on the distribution of the $X_j$ and $Y_j$.

\begin{paragraph}{Definition}
    A family of probability measures $\{ \mathcal{L}^N_{\mu,\sigma} \colon \mu,\sigma\in\mathbb{R}^n\}$ is said to be \emph{admissible} if it satisfies the following properties:
    \begin{enumerate}
        \item The mean and covariance of a random vector $X$ under $\mathcal{L}_{\mu,\sigma}^N$ are $\mu$ and $\text{diag}(\sigma^2)$ respectively. Here, $\text{diag}(\sigma^2)$ denotes the $N\times N$ diagonal matrix with diagonal entries $\sigma_1^2,\ldots,\sigma^2_N$.
        \item For any fixed $\mu\in\mathbb{R}^N$ there exists a constant $C$ such that the fourth moments of a random vector $X$ following $\mathcal{L}^N_{\mu,\sigma}$ remains bounded as $\sigma$ ranges over a given closed bounded subset of $(0,\infty)^N$.
    \end{enumerate}
\end{paragraph}

In the following we will assume that the random vectors $X$ and $Y$ follow an admissible distribution. We note that there is a wide range of families of laws that are admissible. In fact, any family with at least four moments whose moments depend continuously on the parameters $\sigma$ will do, such as the Gaussian, exponential or uniform laws. Thus, imposing admissibility according to our definition is a very mild assumption.

The $L^2$ law of large numbers (see chapter 2 of \cite{durrett_probability_2010}) we have the following: for any closed and bounded set $K\subset (0,\infty)^N$ there exists $C_K>0$ such that for all $\sigma\in K$ we have 
\begin{align}
    \label{eq:l2lln}
        \mathbb{E}\left|\frac1N \sum_{k=1}^N X_k^2 - \frac1N\sum_{k=1}^N \sigma_k^2 \right|^2 \leq \frac{C_K}N
\end{align}
and an analogous bound for the $Y$ (recall that the $X$ and $Y$ follow the same law).

\paragraph{Proposition} Under the assumption of admissibility, for any closed and bounded set $K\subset (0,\infty)^N$ there exists $C_K>0$ such that for all $\sigma\in K$ we have 
\begin{align}
\label{eq:boundCosHatCos}
    \mathbb{E} \left| \cos(X,Y) - \widehat{\cos}(X,Y) \right|^2
    \leq \frac{C_K} N .
\end{align}

Proof. In the following let $W_N=\sqrt{\frac1N \sum_{k=1}^N X_{k}^{2}}$ and $Z_N=\sqrt{\frac1N \sum_{k=1}^{N} Y_{k}^{2}}$. For any integrable random variable $\Psi$ denote by $\overline\Psi$ its expectation: $\overline \Psi = \mathbb{E} \Psi$. We will also write $\Gamma_N= \frac1N \sum_{k=1}^N X_k Y_k$. 
Then
\begin{align}
    \mathbb{E}\left| \cos(X,Y) - \widehat{\cos}(X,Y) \right|^2 & = \frac1{\overline {W_N} \overline {Z_N}} \mathbb{E} \left| \frac{\Gamma_N}{W_N Z_N} \left(W_N Z_N - \overline{W_N} \overline{Z_N} \right) \right|^2\leq \frac1{\overline {W_N} \overline {Z_N}} \mathbb{E} \left| W_N Z_N - \overline{W_N} \overline{Z_N} \right|^2
\end{align}
Here, the second inequality is due to the fact that  $\Gamma_N \leq W_N Z_N$, by the Cauchy-Schwarz inequality. But now the result follows from the $L^2$-law of large numbers \eqref{eq:l2lln}. 

Note that the values for the constants $C_K$ in \eqref{eq:l2lln} and \eqref{eq:boundCosHatCos} are not the same, but this is irrelevant for our argument: we only need to have uniform control of the convergence over compact sets.

It now follows from standard arguments that 
\begin{align}
\label{eq:VariancesClose}
    \left| \text{var } \cos(X,Y) - \text{var } \widehat{\cos} (X,Y)\right| \leq \frac {C_K}N
\end{align}
for another constant $C_K$, as $\sigma$ varies over a fixed closed and bounded set $K$.

From now on, let us emphasise the dependence on the variances and write, for $\lambda_j=\sigma_j^2$,
\begin{align}
    \phi(\lambda_1,...,\lambda_N) = \text{var}\cos(X,Y),\quad \widehat\phi(\lambda_1,...,\lambda_N) = \text{var }\widehat\cos(X,Y).
\end{align}
In this notation, we have shown that for every closed and bounded subset $K$ of $(0,\infty)^N$, there exists $C_K>0$ such that $|\phi(\lambda)-\widehat\phi(\lambda)|<\frac{C_K}N$ whenever $\lambda\in K$.

Note that $\mathbb{E} X_k^2=\mathbb{E} Y_k^2 = u_k + \mu_k^2$. Hence, for $u\in K_N$,
\begin{align}
\label{eq:FormulaPhiHatKN}
    \widehat\phi (u) & = \frac{\text{var}(\sum_{k=1}^N X_k Y_K)}{(\sum_{k=1}^N u_k+\mu_k^2)^2} = \frac1{(1 + \|\mu\|_2^2)^2}\ \left({\sum_{k=1}^N(u_k+\mu_k^2)^2 - \|\mu\|_4^4 } \right)
\end{align}
From \eqref{eq:FormulaPhiHatKN} it follows immediately that $\widehat\phi$ is strictly convex on the compact and convex set $K_N$, which implies that $\widehat\phi$ has a unique minimiser on $K_N$.



\subsection{Approximation of the norm of multivariate vector}\label{normMultivar}
A key step in the approximation of the distribution of cosine similarity for Case 2 and Case 3 is calculating the distribution of the norm of a vector $X$, $\abs{X}$. As before, let $\mathbb{E}[X_{i}] = \mu_{i}$, and $Var(X_{i}) = \sigma_{i}^{2}$. It turns out to be easier to manipulate the square of the norm, $\abs{X}^{2}$. By definition: 
\begin{equation}
    \abs{X}^{2} = \sum_{i} X_{i}^{2}
\end{equation}
The moments are then given by the following. Note that $Var(X_{i}) = \mathbb{E}[X_{i}^{2}] - \mathbb{E}[X_{i}]^{2}$.
\begin{align*}
    \mathbb{E}[\abs{X}^{2}] &= \mathbb{E}[\sum_{i}X_{i}^{2}] = \sum_{i} \mathbb{E} [X_{i}^{2}] \\
                            &= \sum_{i} (Var(X_{i}) + \mathbb{E}[X_{i}]^{2}) 
\end{align*}
\begin{equation}
    \mathbb{E}[\abs{X}^{2}] = \sum_{i} \sigma_{i}^{2}  + \mu_{i}^{2} 
\end{equation}
\begin{align}
    Var(\abs{X}^{2}) &= \mathbb{E}[\sum_{i}(X_{i}^{2})\sum_{j}(X_{j}^{2})] - (\sum_{i} \sigma_{i}^{2} + \mu_{i}^{2})^{2} \\
    &= \sum_{i}\sum_{j} \mathbb{E}[X_{i}^{2}X_{j}^{2}] - (\sum_{i} \sigma_{i}^{2} + \mu_{i}^{2})^{2} \\
    &= \sum_{i} \mathbb{E}[X_{i}^{4}] + \sum_{i}\sum_{j \neq i} \mathbb{E}[X_{i}^{2}]\mathbb{E}[X_{j}^{2}] - (\sum_{i} \sigma_{i}^{2} + \mu_{i}^{2})^{2} \\
    &= \sum_{i} \mathbb{E}[X_{i}^{4}] - \sum_{i} (\sigma_{i}^{2} + \mu_{i}^{2})^{2}
\end{align}

For a normal distribution, $X \sim N(\mu, \Sigma)$ (see Appendix \ref{apdx:normMultinorm}):
\begin{equation}
Var(\abs{X}^{2}) = \sum_{k} 2\sigma_{k}^{2}(\sigma_{k}^{2} + 2 \mu_{k}^{2})
\end{equation}

The mean and variance of the individual components $X_{i}^{2}$ is given by:
\begin{align}
    \mathbb{E}[X_{i}^{2}] \equiv \nu_{i} &= \mu_{i}^{2} + \sigma_{i}^{2} \\
    Var[X_{i}^{2}] &= 2\sigma_{i}^{2}(\sigma_{i}^{2} + 2 \mu_{i}^{2})
\end{align}

Let $s_{n}^{2} = \sum_{i=1}^{n} Var[X_{i}]$.

Suppose Lindeberg's condition holds:
\begin{equation}
    \lim_{n\rightarrow \infty} \frac{1}{s_{n}^{2}} \sum_{k=1}^{n} \mathbb{E} [(X_{k} - \nu_{k})^{2}\textbf{1}_{\{ \abs{X_{k} - \nu_{k}} > \epsilon s_{n}\}}] = 0
\end{equation}

Then it follows from the Central Limit Theorem that:
\begin{equation}
    \lim_{n\rightarrow \infty} \sum_{k=1}^{n} X_{k}^{2} \xrightarrow{d} N(\sum_{k=1}^{n} \nu_{k}, s_{n}^{2})
\end{equation}

The Lindeberg condition effectively constrains the contribution of any individual random variable to the variance to be arbitrarily small for sufficiently large values of n. A constraint like Lindeberg's on the variances is necessary for the squared norm to converge in distribution to a normal distribution, but the mean and variances are exact even in the absence of the constraint on the component variances. 

Concerning the rate of convergence, it useful to consider the ratio of the standard deviation of $\abs{X}^{2}$ to the mean (or the ratio of the variance to the square of the mean):
\begin{align}
    \frac{Var(\abs{X}^{2})}{\mathbb{E}(\abs{X}^{2})^{2}} &= \frac{\sum_{k} 2\sigma_{k}^{2}(\sigma_{k}^{2} + 2\mu_{k}^{2})}{\left(\sum_{k} \sigma_{k}^{2} + \mu_{k}^{2}\right)^{2}} \\
    &= \frac{\sum_{k} 2\sigma_{k}^{2}(\sigma_{k}^{2} + 2\mu_{k}^{2})}{\sum_{k}\sum_{j} (\sigma_{k}^{2} + \mu_{k}^{2})(\sigma_{j}^{2} + \mu_{j}^{2})}
\end{align}

Considering the reciprocal, the square of the mean divided by the variance:
\begin{align*}
    \frac{\mathbb{E}(\abs{X}^{2})^{2}}{Var(\abs{X}^{2})} &= \frac{\sum_{k} (\sigma_{k}^{4} + 2\sigma_{k}^{2}\mu_{k}^{2} + \mu_{k}^{4})}{\sum_{k} 2(\sigma_{k}^{4} + 2\sigma_{k}^{2}\mu_{k}^{2})} + \frac{\sum_{k}\sum_{j\neq k} (\sigma_{k}^{2} + \mu_{k}^{2})(\sigma_{j}^{2} + \mu_{j}^{2})}{\sum_{k} 2(\sigma_{k}^{4} + 2\sigma_{k}^{2}\mu_{k}^{2})} \\
    &= \frac{1}{2} + \frac{\sum_{k} \mu_{k}^{4}}{\sum_{k} 2(\sigma_{k}^{4} + 2\sigma_{k}^{2}\mu_{k}^{2})} + \frac{\sum_{k}\sum_{j\neq k} (\sigma_{k}^{2} + \mu_{k}^{2})(\sigma_{j}^{2} + \mu_{j}^{2})}{\sum_{k} 2(\sigma_{k}^{4} + 2\sigma_{k}^{2}\mu_{k}^{2})}
\end{align*}

Without some constraints on the values of $\sigma_{i}$ and $\mu_{i}$, it is difficult to characterize this ratio. When $\mathbb{E}[X_{i}] = 0 \forall i$, the ratio becomes:
\begin{equation}
    \frac{\mathbb{E}(\abs{X}^{2})^{2}}{Var(\abs{X}^{2})} = \frac{1}{2} + \frac{\sum_{k}\sum_{j \neq k} \sigma_{k}^{2}\sigma_{j}^{2}}{\sum_{k} 2 \sigma_{k}^{4}}
\end{equation}

As an illustration, consider the isotropic case, where $\sigma_{k} = \sigma_{j} \forall j,k$. The ratio then tends to infinity as $n \rightarrow \infty$, indicating that the ratio of the standard deviation to the mean of $\abs{X}^{2}$ converges to 0.
\begin{equation}
    \frac{\mathbb{E}(\abs{X}^{2})^{2}}{Var(\abs{X}^{2})} = \frac{N}{2}
\end{equation}

As the ratio of the variance to the square of the mean of $\abs{X}$ tends to zero, the approximation that $\abs{X} \approx \mathbb{E}[\abs{X}]$ becomes more accurate, making the derivation of the moments of $\cos$ more accurate. 

\clearpage
\section*{Appendix}

\subsection*{Notes}

Here I list helpful relationships that were used in the work. For readability, I switch between notations: $e^{x} = \exp{(x)}$. 

\begin{enumerate}

\item Transformation of a random variable: for a probability density function for a random variable $Z \sim f_{Z}(z)$ with CDF $F_{Z}(z)$, then the probability distribution for a scaled random variable $c Z$ for some $c \in \mathbb{R}$ is given by the following: 

\begin{equation}
F_{cZ}(z) = Pr[cZ < z] = Pr[Z < \frac{z}{c}] = F_{Z}(\frac{z}{c})
\end{equation}
\begin{equation}
f_{cZ} = \frac{d}{dz}[F_{cZ}(z)] = \frac{d}{dz}[F_{Z}(\frac{z}{c})] = \frac{1}{c}f_{Z}(\frac{z}{c})
\end{equation}

\item Variance of the sum of independent random variables is the sum of the variances. Let X, Y be independent random variables. We make use of $\mathbb{E}[XY] = \mathbb{E}[X]\mathbb{E}[Y]$ for independent variables. 
\begin{align}
    Var(X+Y) &= \mathbb{E}[(X+Y)^{2}] - \mathbb{E}[(X+Y)]^{2} \\
         &= \mathbb{E}[X^{2} + 2XY + Y^{2}] - (\mathbb{E}[X] + \mathbb{E}[Y])^{2} \\
         &= \mathbb{E}[X^{2}] + 2\mathbb{E}[XY] + \mathbb{E}[Y^{2}] - \mathbb{E}[X]^2 - \mathbb{E}[Y]^2 - 2\mathbb{E}[X]\mathbb{E}[Y] \\
         &= Var(X) + Var(Y)
\end{align}

\item Variance of the product of independent random variables:
\begin{align}
    Var(XY) &= \mathbb{E}[X^{2}Y^{2}] - (\mathbb{E}[XY])^{2} \\
        &= \mathbb{E}[X^{2}]\mathbb{E}[Y^{2}] - (\mathbb{E}[X])^{2}(\mathbb{E}[Y])^{2} \\
        &= Var(X)Var(Y) + Var(X)\mathbb{E}[Y]^{2} + Var(Y)\mathbb{E}[X]^{2}
\end{align}
It then follows that, if $\mathbb{E}[X] = \mathbb{E}[Y] = 0$:
\begin{equation}
    Var(XY) = Var(X)Var(Y)
\end{equation}

\item Characteristic function of a normal random variable. Let $X \sim N(\mu, \sigma^{2})$.  
\begin{align}
    \phi_{X}(t) &= \mathbb{E}[e^{itx}] = \int_{R} e^{itx} \frac{1}{\sigma \sqrt{2\pi}} e^{-\frac{(x-\mu)^{2}}{2\sigma^{2}}} dx \\
                &= \frac{1}{\sigma \sqrt{2\pi}} \int_{R} dx \exp{[(itx - \frac{1}{2\sigma^{2}}(x^{2} - 2x\mu + \mu^{2})]} \\
                &= \frac{1}{\sigma \sqrt{2\pi}} \exp{[-\frac{1}{2\sigma^{2}}(\mu^{2} - (\mu^{2}+\sigma^{2}it)^{2})]}\int_{R} dx \exp{[-\frac{1}{2\sigma^{2}}(x - (\mu + \sigma^{2}it))^{2}]} \\
                &= \frac{1}{\sigma \sqrt{2\pi}} \exp{[-\frac{1}{2\sigma^{2}}(- \sigma^{4}t^{2} + 2\mu\sigma^{2}it)]}\sqrt{2\pi\sigma^{2}} \\
                &= \exp{[i\mu t - \frac{\sigma^{2}t^{2}}{2}]} \label{charNorm}
\end{align}

\item Sum of normally distributed random variables is normal: 
This is easiest to show with characteristic functions. The characteristic function $\phi_{X}$ for $X \sim N(\mu, \sigma^2)$ is:
\begin{equation}
    \phi_{X}(t) = \exp{[i\mu t - \frac{\sigma^{2}t^{2}}{2}]}
\end{equation}
Let $X_{j}$, $j = 1:n$ be a sequence of independent random variables, where $X_{j} \sim N(\mu_{j}, \sigma_{j}^2)$. For $\alpha_{j} \in \mathbb{R}$ for $j = 1:n$, let $S = \sum_{j} \alpha_{j}X_{j}$. 
\begin{align}
    \phi_{S}(t) &= \prod_{j} \phi_{X_{j}}(\alpha_{j}t) \\
                &= \prod_{j} \exp[i\mu_{j}\alpha_{j}t - \frac{\sigma_{j}^{2}\alpha_{j}^{2} t^{2}}{2}] \\
                &= \exp[(\sum_{j} i\mu_{j}\alpha_{j}t) - \frac{1}{2}(\sum_{j} \sigma_{j}^{2}\alpha_{j}^{2})t^{2}]
\end{align}
Thus, $S \sim N(\sum_{j} \alpha_{j}\mu_{j}, \sum_{j}\alpha_{j}^{2} \sigma_{j}^{2})$. 

\item The square of a centered normally distributed random variable is gamma distributed:
Let $X \sim N(0, \sigma^{2})$. Define $g(X) = X^{2}$. We apply the Law of the Unconscious Statistician; the characteristic function $\phi_{X^{2}}$ of $X^{2}$ is given as follows. Define a: $a \equiv \frac{1}{2\sigma^{2}}(1-2\sigma^{2}it)$
\begin{align}
    \phi_{X^{2}}(t) &= \mathbb{E}[e^{itx^{2}}] \\
                 &= \int_{R} e^{itx^{2}} \frac{1}{\sigma \sqrt{2\pi}} e^{-\frac{x^{2}}{2\sigma^{2}}} dx \\
                 &= \frac{1}{\sigma\sqrt{2\pi}} \int_{R} e^{-ax^{2}} \\
                 &= \frac{1}{\sigma\sqrt{2\pi}}\sqrt{{\frac{\pi}{a}}} \\
                 &= \frac{1}{\sigma\sqrt{2a}} \\
                 &= (1 - 2\sigma^{2}it)^{-1/2}
\end{align}
$\phi_{X^{2}}$ is the characteristic function of a gamma distributed random variable with scale parameter $\theta = 2\sigma^{2}$ and shape parameter $k = \frac{1}{2}$, i.e. $X^{2} \sim \Gamma(k=\frac{1}{2}, \theta = 2\sigma^{2})$. 

\item Characteristic function of the square of a normal random variable.
Let $X \sim N(\mu, \sigma^{2})$. Define $g(X) = X^{2}$. The characteristic function $\phi_{X^{2}}$ of $X^{2}$ is given as follows, where the third step follows from completing the square:
\begin{align}
\phi_{X^{2}}(t) &= \mathbb{E}[e^{itx^{2}}] \\
                &= \int_{R} e^{itx^{2}} \frac{1}{\sigma \sqrt{2\pi}} e^{-\frac{(x-\mu)^{2}}{2\sigma^{2}}} dx \\
                &= \frac{1}{\sigma \sqrt{2\pi}} \int_{R} \exp{\frac{it\mu^{2}}{(1 - 2it\sigma^{2})}} \exp{[(it - \frac{1}{2\sigma^{2}})(x-\frac{\mu}{1 - 2it\sigma^{2}} )^{2}}] dx \\
                &= \frac{1}{\sigma \sqrt{2\pi}} e^{\frac{it\mu^{2}}{(1 - 2it\sigma^{2})}} \int_{R} \exp{[(it - \frac{1}{2\sigma^{2}})(x-\frac{\mu}{1 - 2it\sigma^{2}} )^{2}}] dx \\
                &= \frac{1}{\sigma \sqrt{2\pi}} e^{\frac{it\mu^{2}}{(1 - 2it\sigma^{2})}} \sqrt{\frac{2\sigma^{2}\pi}{1 - 2\sigma^{2}}} \\ 
                &= \exp{[\frac{it\mu^{2}}{1 - 2it\sigma^{2}}]} \sqrt{\frac{1}{1 - 2\sigma^{2}it}} \label{eqn:squareNorm} 
\end{align}

\item Characteristic function of the product of two independent normal random variables:
Let $X \sim N(\mu_{1}, \sigma_{1}^{2})$, $Y \sim N(\mu_{2}, \sigma_{2}^{2})$. Define $W = XY$. The characteristic function for W is given by:

\begin{align}
    \phi_{W}(t) &= \mathbb{E}[e^{itW}] = \mathbb{E}[e^{itXY}] \\
                &= E_{Y}[ E_{X} [ e^{itXY} | Y]]
\end{align}
Where this last statement follows from the law of total probability. Substituting $(tY)$ in for $t$ in the characteristic function of a normal (Equation \ref{charNorm}) solves the first expectation over $X$. For conciseness, define $G \equiv (1 + \sigma_{1}^{2}\sigma_{2}^{2}t^{2})$

\begin{align*}
    \phi_{W}(t) &= E_{Y}[ \exp(i\mu_{1}tY - \frac{\sigma_{1}^{2}t^{2}Y^{2}}{2})] \\
                &= \frac{1}{\sigma_{2}\sqrt{2\pi}}\int_{R} dy \exp\left( i\mu_{1}ty - \frac{\sigma_{1}^{2}t^{2}y^{2}}{2}\right) e^{-\frac{(y - \mu_{2})^{2}}{2\sigma_{2}^{2}}} \\
                &= \frac{1}{\sigma_{2}\sqrt{2\pi}}\int_{R} dy \exp\left(-\frac{G}{2\sigma_{2}^{2}}\left(y^{2} - \frac{2\sigma_{2}^{2}i\mu_{1}t + \mu_{2}}{G}y + \frac{\mu_{2}^{2}}{G}\right)\right) \\
                &= \frac{1}{\sigma_{2}\sqrt{2\pi}}\int_{R} dy \exp\left(-\frac{G}{2\sigma_{2}^{2}} \left(\left(y - \frac{i\mu_{1}t\sigma_{2}^{2} + \mu_{2}}{G}\right)^{2} + \frac{\mu_{2}^{2}}{G} - \left(\frac{i\mu_{1}t\sigma_{2}^{2} + \mu_{2}}{G}\right)^{2}\right)\right) \\
                &= \frac{1}{\sigma_{2}\sqrt{2\pi}} \exp\left(-\frac{G}{2\sigma_{2}^{2}}\left(\frac{\mu_{2}^{2}}{G} - \left(\frac{i\mu_{1}t\sigma_{2}^{2} + \mu_{2}}{G}\right)^{2}\right)\right) \\ 
                &\int_{R} dy \exp\left(-\frac{G}{2\sigma_{2}^{2}}\left(y - \frac{i\mu_{1}t\sigma_{2}^{2} + \mu_{2}}{G}\right)^{2}\right) \\
                &= \exp\left( -\frac{\mu_{2}^{2}}{2\sigma_{2}^{2}} + \frac{(i\mu_{1}t\sigma_{2}^{2} + \mu_{2})^{2}}{2\sigma_{2}^{2}G}\right) \frac{1}{\sigma_{2}\sqrt{2\pi}} \sqrt{\frac{2\pi\sigma_{2}^{2}}{1 + \sigma_{1}^{2}\sigma_{2}^{2} t^{2}}} \\
                &= (1 + \sigma_{1}^{2}\sigma_{2}^{2}t^{2})^{-1/2} \exp\left(\frac{-\mu_{2}^{2}(1 + \sigma_{1}^{2}\sigma_{2}^{2}t^{2}) + \mu_{2}^{2} + 2i\mu_{1}\mu_{2}\sigma_{2}^{2}t - \mu_{1}\sigma_{2}^{4}t^{2}}{2\sigma_{2}^{2}(1 + \sigma_{1}^{2}\sigma_{2}^{2}t^{2})}\right) 
\end{align*}
\begin{equation}
        \phi_{W}(t) = (1 + \sigma_{1}^{2}\sigma_{2}^{2}t^{2})^{-1/2} \exp\left(\frac{i \mu_{1}\mu_{2}t}{(1 + \sigma_{1}^{2}\sigma_{2}^{2}t^{2})} - \frac{\left(\mu_{1}^{2}\sigma_{2}^{2} + \mu_{2}^{2}\sigma_{1}^{2}\right)t^{2}}{2(1 + \sigma_{1}^{2}\sigma_{2}^{2}t^{2})}\right) \label{eqn:charProd}
\end{equation}

\item Norm (Euclidean length) of a multivariate normal random variable. \label{apdx:normMultinorm}
Let $X_{i} \sim N(\mu_{i}, \sigma_{i}^{2})$, i.e. X is a multivariate normal with mean $\boldsymbol{\mu} = \mu_{i}$ and diagonal covariance matrix $\Sigma$, where $\Sigma_{ii} \equiv \sigma_{i}^{2}$ and $\Sigma_{ij} = 0$ for $i \neq j$. Let $Q = |X|^{2}$ (the "Q" is for "quadratic"). 
\begin{equation}
Q \equiv |X|^{2} = \sum_{i} X_{i}^{2}
\end{equation}
Per Equation \ref{eqn:squareNorm}, we have the characteristic function of Q:
\begin{align}
    \phi_{Q}(t) &= \prod_{i} \phi_{X_{i}^{2}} \\
        &= \prod_{i} \exp \left(\frac{it\mu_{i}^{2}}{1 - 2it\sigma_{i}^{2}}\right) \sqrt{\frac{1}{1 - 2it\sigma_{i}^{2}}} 
\end{align}
It is useful to write the derivatives of $\phi_{X_{i}^{2}}(t)$, where for conciseness, $B_{i} \equiv (1 - 2it\sigma_{i}^{2})$
\begin{equation}
    \frac{d}{dt}\phi_{X_{i}^{2}}(t) = \exp \left(\frac{it\mu_{i}^{2}}{1 - 2it\sigma_{i}^{2}}\right) (1 - 2\sigma_{i}^{2}it)^{-1/2}
            \left(\frac{i\sigma_{i}^{2}}{1 - 2it\sigma_{i}^{2}} - \frac{2\mu_{i}^{2} \sigma_{i}^{2} t}{(1 - 2it\sigma_{i}^{2})^{2}} + \frac{i\mu_{i}^{2}}{1 - 2it\sigma_{i}^{2}}\right)
\end{equation}
\begin{equation}
    \frac{d^{2}}{dt^{2}} \phi_{X_{i}^{2}}(t) = \exp \left(\frac{it\mu_{i}^{2}}{1 - 2it\sigma_{i}^{2}}\right) (B_{i})^{-1/2}
            \left( \frac{-2\sigma_{i}^{4}}{B_{i}^{2}} - \frac{4\mu_{i}^{2}\sigma_{i}^{2}}{B_{i}^{2}} - \frac{8i\mu_{i}^{2}\sigma_{i}^{4}t}{B_{i}^{3}} + \left[\frac{i\sigma_{i}^{2}}{B_{i}} - \frac{2\mu_{i}^{2} \sigma_{i}^{2} t}{B_{i}^{2}} + \frac{i\mu_{i}^{2}}{B_{i}} \right]^{2}   \right)
\end{equation}
It follows then that:
\begin{align}
    \mathbb{E}[X_{i}^{2}] &= \mu_{i}^{2} + \sigma_{i}^{2} \\
    \mathbb{E}[(X_{i}^{2})^{2}] &= 3 \sigma_{i}^{4} + 6 \mu_{i}^{2}\sigma_{i}^{2} + \mu_{i}^{4} \\ 
    Var(X_{i}^{2}) &= 2 \sigma_{i}^{2} (\sigma_{i}^{2} + 2 \mu_{i}^{2})
\end{align}
\begin{enumerate}
    \item $\mathbb{E}[Q] = \mathbb{E}[X^{2}]$ - the expectation of the square of the norm of X. For conciseness, define $B_{k} \equiv (1 - 2it \sigma_{k}^{2})$
    \begin{align*}
        \mathbb{E}[Q] &= \frac{1}{i} \frac{d}{dt} \phi_{Q} |_{t = 0} \\
             &= \sum_{k} \left( e^{\frac{it\mu_{k}^{2}}{B_{k}}} B_{k}^{-1/2}\left(\sigma_{k}^{2} B_{k}^{-1} + 2i \mu_{k}^{2}\sigma_{k}^{2}t + B_{k}^{-1}\mu_{k}^{2}\right) \prod_{j \neq k} \exp \left(\frac{it\mu_{j}^{2}}{B_{j}}\right) \sqrt{\frac{1}{B_{j}}} \right)_{t=0} \\
             &= \sum_{k} (\sigma_{k}^{2} + \mu_{k}^{2})
    \end{align*}
    \begin{equation}
        \mathbb{E}[Q] = \mathbb{E}[|X|^{2}] = \sum_{k}(\sigma_{k}^{2} + \mu_{k}^{2}) 
        \label{normLengthVar}
    \end{equation}
    \item $\mathbb{E}[Q^{2}] = \mathbb{E}[X^{4}]$ - the second moment of Q, the square of the norm of X. 
    \begin{align*}
        \mathbb{E}[Q^{2}] &= (-1)\frac{d^{2}}{dt^{2}} \phi_{Q}|_{t = 0} \\
                 &= (-1)\left[\sum_{k} \left( (\prod_{i \neq k} \phi_{X_{i}^{2}}) \frac{d^{2}}{dt^{2}}\phi_{X_{k}^{2}}  \right) + \sum_{j}\sum_{k\neq j} (\prod_{i \neq j,k} \phi_{X_{i}^{2}}) \frac{d}{dt}\phi_{X_{j}^{2}} \frac{d}{dt}\phi_{X_{k}^{2}}\right]  \\
                 &= \sum_{k} (3\sigma_{k}^{4} + 6\mu_{k}^{2} \sigma_{k}^{2} + \mu_{k}^{4}) + \sum_{j}\sum_{k \neq j} (\mu_{j}^{2} + \sigma_{j}^{2})(\mu_{k}^{2} + \sigma_{k}^{2})
    \end{align*}
    The variance is then:
    \begin{align*}
        Var(Q) &= \mathbb{E}[Q^{2}] - \mathbb{E}[Q]^{2} \\
               &= \sum_{k} (3\sigma_{k}^{4} + 6\mu_{k}^{2} \sigma_{k}^{2} + \mu_{k}^{4}) + \sum_{j}\sum_{k \neq j} (\mu_{j}^{2} + \sigma_{j}^{2})(\mu_{k}^{2} + \sigma_{k}^{2}) - \sum_{j}\sum_{k}(\sigma_{j}^{2} + \mu_{j}^{2})(\sigma_{k}^{2} + \mu_{k}^{2})) \\
               &= \sum_{k} (3\sigma_{k}^{4} + 6\mu_{k}^{2} \sigma_{k}^{2} + \mu_{k}^{4}) - \sum_{k}(\sigma_{k}^{2} + \mu_{k}^{2})^{2}
    \end{align*}
    \begin{equation}
        Var(Q) = Var(\abs{X}^{2}) = \sum_{k} 2 \sigma_{k}^{2} (\sigma_{k}^{2} + 2\mu_{k}^{2})
    \end{equation}
    \item $\mathbb{E}[|X|]$, the expectation of the (Euclidean) norm of $X$, turns out to be surprisingly difficult to calculate. There is a useful relationship presented as Theorem 3.2b.5 in \cite{mathai_quadratic_1992}. First, we put an upper bound on the expectation. Note that $f(x) = x^{2}$ is convex, as the second derivative, $f''(x) = 2$, is strictly positive. Then, by Jensen's inequality:
    \begin{align*}
        f(\mathbb{E}[|X|]) &\leq \mathbb{E}[f(|X|)] \\ 
        \mathbb{E}[|X|]^{2} &\leq \mathbb{E}[|X|^{2}] \\ 
                   &\leq \mathbb{E}[\sum_{j} x_{j}^{2}] \\ 
                   &\leq \sum_{j} \mathbb{E}[x_{j}^{2}] = \sum_{j} \mu_{j}^{2} + \sigma_{j}^{2}
    \end{align*}
    \begin{equation}
        \mathbb{E}[|X|] \leq \sqrt{\sum_{j} \mu_{j}^{2} + \sigma_{j}^{2}}
        \label{normLengthMean}
    \end{equation}
    
\end{enumerate}

\end{enumerate}

\subsection*{Author Contributions}

I.S. conceptualized the study. I.S. and J.O. did the derivations with input from F.A.A. and P.S. I.S. designed and implemented the simulations. I.S. wrote the manuscript with input from all authors. B.H.K. supervised the study and refined the analysis. 

\printbibliography

\end{document}